\documentclass[aps,amssymb,amsmath,reprint,superscriptaddress,showpacs]{revtex4-1}
\usepackage{bm}
\usepackage{graphicx}
\usepackage{color}

\usepackage{mathtools}
\usepackage{amsfonts}
\usepackage{textcomp} 
\usepackage{microtype} 
\usepackage{hyperref}
\hypersetup{colorlinks=true}
\usepackage[all]{hypcap} 

\newcommand{\akd}{a^{\dagger}_{k}}
\newcommand{\ak}{a^{\phantom{\dagger}}_{k}}
\newcommand{\aqd}{a^{\dagger}_{q}}
\newcommand{\aq}{a^{\phantom{\dagger}}_{q}}
\newcommand{\annd}{a^{\dagger}_{n}}
\newcommand{\an}{a^{\phantom{\dagger}}_{n}}
\newcommand{\w}{\omega}

\newcommand{\ra}{\rangle}

\newcommand{\mr}[1]{{{\mathrm{#1}}}}

\newcommand{\s}{\sigma}

\newcommand{\ad}{a^{\dagger}}
\newcommand{\apd}{a^{\phantom{\dagger}}}

\begin{document}

\title{
Anatomy of quantum critical wave functions in dissipative impurity problems
}

\author{Zach Blunden-Codd} 
\affiliation{Photon Science Institute \& School of
Physics and Astronomy, University of Manchester, Oxford Road, Manchester M13
9PL, United Kingdom}
\affiliation{Department of Physics, Imperial College London, London
SW7 2AZ, United Kingdom}
\author{Soumya Bera}
\affiliation{Max-Planck-Institut fuer Physik komplexer Systeme, 01187 Dresden,
Germany}
\author{Benedikt Bruognolo}
\affiliation{Physics Department, Arnold Sommerfeld Center for Theoretical
Physics, and Center for NanoScience, Ludwig-Maximilians-Universi\"at, 
Theresienstrasse 37, 80333 M\"unchen, Germany}
\affiliation{Max-Planck-Institut ur Quantenoptik, Hans-Kopfermann-Str. 1, 85748
Garching, Germany}
\author{Nils-Oliver Linden}
\affiliation{Physics Department, Arnold Sommerfeld Center for Theoretical
Physics, and Center for NanoScience, Ludwig-Maximilians-Universi\"at, 
Theresienstrasse 37, 80333 M\"unchen, Germany}
\author{Alex W. Chin}
\affiliation{Theory of Condensed Matter Group, University of Cambridge,
J. J. Thomson Avenue, Cambridge, CB3 0HE, United Kingdom}
\author{Jan von Delft}
\affiliation{Physics Department, Arnold Sommerfeld Center for Theoretical
Physics, and Center for NanoScience, Ludwig-Maximilians-Universi\"at, 
Theresienstrasse 37, 80333 M\"unchen, Germany}
\author{Ahsan Nazir}
\affiliation{Photon Science Institute \& School of Physics and Astronomy,
University of Manchester, Oxford Road, Manchester M13 9PL,
United Kingdom}
\author{Serge Florens}
\affiliation{Institut N\'{e}el, CNRS and Universit\'e Grenoble Alpes, F-38042
Grenoble, France}

\begin{abstract}
Quantum phase transitions reflect singular changes taking place in a many-body 
ground state, however, computing and analyzing large-scale critical wave functions 
constitutes a formidable challenge.
New physical insights into the sub-Ohmic spin-boson model are provided by the 
coherent state expansion (CSE), which represents the wave function by a linear 
combination of classically displaced configurations. We find that the distribution 
of low-energy displacements displays an emergent symmetry in the absence of 
spontaneous symmetry breaking, while experiencing strong fluctuations 
of the order parameter near the quantum critical point. Quantum criticality
provides two strong fingerprints in critical low-energy modes: an algebraic
decay of the average displacement and a constant universal average squeezing 
amplitude. These observations, confirmed by extensive variational matrix product 
states (VMPS) simulations and field theory arguments, offer precious clues into 
the microscopics of critical many-body states in quantum impurity models.

\end{abstract}

\date{\today}

\maketitle

\section{Introduction}
The understanding of critical phenomena in classical mechanics owes a great deal to the
spatial representation of critical states, whereby the order parameter
experiences statistical fluctuations on all length scales, due to a diverging correlation
length~\cite{Ma,Amit} at the critical temperature. This scale invariance property was the
starting point for one of the most powerful tools in theoretical physics, the
renormalization group, which allowed rationalization of classical criticality in terms of
trajectories in the space of coupling constants~\cite{Wilson}.
Today, one frontier of research in critical phenomena lies in the 
quantum realm, where criticality may govern some of the most fascinating and 
complex properties found in strongly correlated materials or cold
atoms~\cite{Carr,Sachdev}.
One very fruitful approach is to consider quantum criticality in light 
of an effective classical theory in higher dimensions~\cite{Sachdev},
combining spatial and temporal fluctuations within the path integral formalism.
Quantum phase transitions are then probed through physical response functions
that display a diverging correlation length in space-time. However, this point of view 
does not provide a full picture of the physics at play, especially since quantum 
criticality pertains to a singular change in a many-body ground state. Developing 
wave function-based approaches to strong correlations is indeed a blossoming field, ranging from 
quantum chemistry~\cite{CoupledCluster} to quantum information~\cite{Verstraete,Schollwoeck2011}, 
so that hopes are high that quantum critical states may be rationalized in a simpler way. 

Our aim in this article is to directly study the quantum critical wave function
of a simple toy-model, the sub-Ohmic spin-boson Hamiltonian, and to unveil some
salient fingerprints of criticality in its ground state. In this standard model, 
to be described in further details below, a single quantized spin interacts with a 
continuum of bosonic modes, with a spectrum of coupling constants that vanishes
with a power-law $s<1$ at low energy.
For this purpose, we shall use a combination of two numerically-exact 
wave function-based methods for quantum impurity models, a variational
matrix-product-state approach
(VMPS)~\cite{White1992,Schollwoeck2005,Weichselbaum,Schollwoeck2011}
and the coherent state expansion (CSE)~\cite{Bera1,Bera2,Gheeraert,Snyman2}. 
VMPS is an acronym for the variational matrix-product-state (MPS)
formulation of the density matrix renomalization group (DMRG), which has 
been established as a very powerful and flexible technique also in the context 
of bosonic impurity models~\cite{Weichselbaum,Guo,Bruognolo}, and will be used 
as a reference.  Its all-purpose character makes it hard, however, to rationalize 
the precise content of the wave function in simple physical terms. For this
reason, we implement the CSE variationally, which amounts to expanding
environmental states of the bath onto a discrete set of classical-like 
configurations, namely coherent states of the bosonic states in the bath
(note that an infinite discrete set is enough to ensure completeness
of the coherent state basis~\cite{Cahill}).
Thus, crucial aspects of quantum criticality can be directly inferred by 
reading-off the various superpositions of oscillator displacements that 
parametrize the set of coherent states.

For a given spin orientation of the impurity, we find that the 
distribution of displacements within the CSE wave function displays an emergent
symmetry (between positive and negative values) in the critical domain.
This implies that the average displacement decays to zero for low-energy 
critical modes, with a universal exponent controlled by the dynamical 
susceptibility. This behavior reflects the absence of spontaneous symmetry 
breaking, and the fact that the magnetization order parameter directly couples 
to the bosonic displacement field. Hence the displacements of the oscillators
in the critical many-body wave function vanish in average at low energy.
In addition, the CSE wave function indicates that the distribution of
displacements admits a finite width at the quantum critical point (although
its mean value vanishes algebraically for critical modes, as mentioned above). 
This observation translates physically the wide fluctuations of the order
parameter that take place in the quantum critical regime in absence of
ordering. At the level of the critical wave function, this effects amounts to 
a finite average squeezing amplitude of the quantum critical modes (averaged 
over a logarithmic energy interval), which we show from field theory arguments 
to take a constant universal value.

For the spin-boson model, we demonstrate that both the MPS and CSE methodologies 
converge to the same results, both away from and at the critical point. We find
that the number of coherent states required to capture quantum critical behavior
on a reasonable energy range (at least three decades) is relatively large, of
the order of a hundred. For this reason, recent investigations of the sub-Ohmic
model with variational CSEs using fewer states~\cite{Zhao,Duan} failed to grasp the critical
exponents found in large scale VMPS calculations~\cite{Guo,Bruognolo}.
In contrast to the usual Kondo problem associated with the Ohmic spin-boson model, 
the sub-Ohmic case is indeed governed by two energy scales in its delocalized phase, 
namely the renormalized tunneling amplitude and the mass of a soft bosonic collective 
mode which drives the transition. Capturing the critical softening requires careful 
and extensive numerical simulations, as we shall show by benchmarking the 
VMPS and CSE against each other.

The paper is organized as follows. In Sec.~\ref{Model}, we present the spin-boson
model, its discretization on a Wilson energy mesh, and the variational solution
of its many-body wave function using both MPS and CSE representations. The
wave function obtained by CSE is displayed to guide physical intuition in the
rest of the paper.
Sec.~\ref{Analytical} develops the necessary analytical work that relates the dynamical
critical exponent of the spin susceptibility to two important features of
the wave function: the average displacement of the environmental state and the average 
width (or squeezing amplitude). This allows us to elucidate
the different behaviors of the wave function in both the non-critical 
delocalized phase and at the quantum critical point.
Finally, Sec.~\ref{Comparison} shows numerical results from the 
VMPS and CSE approaches, finding excellent agreement between each other,
as well as with analytical predictions. Appendix~\ref{Appendix} provides details
on a new hierarchical algorithm devised to solve the CSE in a fast and reliable
way.

\section{Ground state wave function of the sub-Ohmic spin-boson model}
\label{Model}

\subsection{Model}

Our study will be based on the spin-boson 
Hamiltonian~\cite{Leggett,Weiss,VojtaReview,Carr,LeHurReview}
with $\Delta$ the quantum Larmor frequency of a two-level system described
by Pauli matrices $\vec{\sigma}$:
\begin{equation}
H = \frac{\Delta}{2} \s_x - \frac{\s_z}{2} 
\sum_k g_k (\akd+\ak) + \sum_k \w_k \akd \ak.
\label{ham}
\end{equation}
The bosonic spectrum assumes a pure power-law with exponent $0\leq s\leq1$
up to a sharp high-energy cutoff $\w_c$ ($\w_c=1$ in all our numerical
computations):
\begin{equation}
J(\w) \equiv \sum_k \pi g_k^2 \delta(\w-\w_k) =
2 \pi \alpha \w_c^{1-s}
\w^s\theta(\w)\theta(\w_c-\w).
\label{bath}
\end{equation}
The Ohmic case ($s=1$) can be realized in the context of 
waveguide-QED~\cite{LeHur,Goldstein,Peropadre,Snyman1}
by coupling a superconducting qubit to a high impedance
transmission line consisting of a uniform Josephson junction 
array. In principle, a precise tailoring of the capacitance
network could allow the sub-Ohmic regime to be realized as well.
In terms of quantum critical phenomena~\cite{Bulla,Tong,Rieger,Anders,Freyn}
the sub-Ohmic model with $0\leq s<1$ presents a continuous
quantum phase transition at a critical coupling $\alpha_c$ between
a localized phase (with $\big<\sigma_z\big>\neq 0$ for $\alpha>\alpha_c$) 
and a symmetric phase (with $\big<\sigma_z\big>=0$ for $\alpha\leq\alpha_c$), 
which will be our focus.

\subsection{Wilson discretization}
The bosonic bath $J(\w)$ is discretized in a logarithmic fashion, using 
a Wilson parameter $\Lambda>1$, first on the highest energy window close to the
cutoff $[\Lambda^{-1}\w_c,\w_c]$, and then iteratively on successive decreasing
energy intervals $[\w_{n+1},\w_n]$ with
$\w_n=\Lambda^{-n}\w_c$~\cite{Bulla,Tong,Guo}.
This leads to the so-called star-Hamiltonian, which involves the direct coupling
of the spin to all bosonic modes (and not to a single site within an extended
bosonic chain):
\begin{equation}
H_\mathrm{star}=\frac{\Delta}{2}\sigma_x
-\frac{1}{2}\sigma_z \sum_{n=0}^{+\infty} \frac{\gamma_n}{\sqrt{\pi}}
[\ad_n+\apd_n]+ \sum_{n=0}^{+\infty} \xi_n \ad_n \apd_n.
\label{star}
\end{equation}
The impurity coupling strength reads 
\begin{equation}
\gamma_n^2=\int_{\w_{n+1}}^{\w_n}\!\!\!\! d\w\, J(\w) =
2\pi\alpha \frac{1-\Lambda^{-(s+1)}}{s+1} \w_c^2 \Lambda^{-n(s+1)},
\label{gamma}
\end{equation}
and the typical energy $\xi_n$ in each Wilson shell is
\begin{equation}
\xi_n = \frac{1}{\gamma_n^2} \int_{\w_{n+1}}^{\w_n}\!\!\!\! d\w\, \w\, J(\w) =
\frac{s+1}{s+2}\frac{1-\Lambda^{-(s+2)}}{1-\Lambda^{-(s+1)}}\w_c
\Lambda^{-n}.
\label{xi}
\end{equation}
Note that the continuum limit is only recovered for $\Lambda\to1$ and an
infinite number of Wilson shells. However, in practice $\Lambda=2$ will be used
in the following, and 50 sites will be used for both the MPS and the CSE
variational calculations. This standard choice of parameters offers a good
compromise between energy resolution and numerical costs, but our techniques
can be pushed in principle to smaller $\Lambda$ values.

\subsection{Variational matrix product states approach} 

One very successful approach that enables direct access to ground-state 
wavefunction of low-dimensional quantum system is the density matrix
renormalization group (DMRG) \cite{White1992,Schollwoeck2005}. Though originally 
developed in the context of one-dimensional real-space systems, the
matrix-product-state formulation of this variational method (VMPS) has been
established as indispensable tool also in the context of quantum impurity models
\cite{Weichselbaum,Schollwoeck2011,Guo,Bruognolo}.

Its application to the spin-boson model works as follows. First, the star
Hamiltonian $H_{\rm star}$ is mapped on a truncated Wilson chain, where the
spin-$\frac{1}{2}$ impurity is coupled to a length-$N$ tight-binding chain model
whose hopping matrix elements decrease exponentially with site number $k$. Next,
one initializes a random MPS for the Wilson chain Hamiltonian,
\begin{equation}
\label{eq:MPS1}
|\psi\rangle = \sum_{\sigma,\mathbf{m}} A^{[\sigma]} A^{[m_0]} A^{[m_1]} \,...
\, A^{[m_{N}]} |\sigma\rangle | \mathbf{m} \rangle\,,
\end{equation}
where $|$$\uparrow$$\rangle, |$$\downarrow$$\rangle$ represents the $\sigma_z$
eigenstates of the impurity  and $ |\mathbf{m}\rangle = |m_0\rangle ...
|m_N\rangle$ describes boson number eigenstates in a truncated Fock basis, i.e.,
$\hat{m}_k |\mathbf{m}\rangle = m_k |\mathbf{m}\rangle$, with $m_k =
0,1,...,d_k-1$. The wave function coefficient is split into a product of tensors
$A^{[...]}$, which are iteratively varied with respect to the energy for finding 
the best approximation for the ground-state wavefunction. If the parameters such as
the bond dimension $D$ and the Fock-space dimension $d_k$ are chosen
appropriately large, the algorithm converges the MPS to a numerically
quasi-exact representation of the ground-state wavefunction.
In practice, we use an optimal boson basis \cite{Guo,Bruognolo} mapping 
the local Fock basis $|m_k\rangle$ to a
smaller, effective bosonic basis $|\tilde{m}_k\rangle$ for efficiency reasons.
Good convergence is ensured for the delocalized phase and at the quantum
critical point for $D=60$, $d_k=100$, and $\tilde{d}_k=16$.

\subsection{Coherent state expansion}

\subsubsection{General methodology}

More recently, an alternative representation of bosonic environmental
wave functions was proposed~\cite{Bera1,Bera2}, based on a simple physical picture of the
energy landscape in terms of classical-like configurations. These are parametrized
by multimode coherent states,
$ |\pm f^{(m)} \ra = e^{\pm
\sum_{k} f_k^{(m)} \left(\akd-\ak\right)} |0\ra $,
with $f_k^{(m)}$ the displacement of mode $k$ for the $m^\mr{th}$ variational
coherent state. Note that the index $k$ labels momentum, while the index
$m=1\ldots M_\mathrm{cs}$ represents an optimal choice of a set of discrete 
coherent states, which embodies a complete basis for an infinite number
of coherent states, $M_\mathrm{cs}\to\infty$~\cite{Cahill}. The expansion 
for the many-body ground state wave function $|\mr{GS}\rangle$ reads:
\begin{equation}
|\mr{GS}\ra = \sum_{m=1}^{M_\mr{cs}}  \Big[ p_m  |f^{(m)} \ra
|\uparrow\rangle + q_m |h^{(m)} \ra |\downarrow\rangle \Big],
\label{eq:GS}
\end{equation}
with the normalization $\big<\mr{GS}|\mr{GS}\big>=1$.
Here, $p_m$ and $q_m$ characterize the weight of the different 
coherent state components within the ground state wave function
for each spin orientation. The discrete sum over the index $m$ can thus be
interpreted as an optimal discretization of the multidimensional 
integral involved in the standard overcomplete
Glauber-Sudarshan representation~\cite{Cahill} in terms of continously
varying displacement functions. We find in practice that the
coherent state representation does not show signs of this overcompleteness 
once the wavefunction is developed on a discrete sum of coherent states,
as in Eq.~(\ref{eq:GS}), and if the number of coherent states $M_\mr{cs}$ 
is typically much less than the number of states in the Hilbert space required 
to capture the ground state (which corresponds to the usual application 
of the method). There is of course a trivial redundancy when reshuffling the
indices $m$ of the set of coherent states for a given solution, but apart 
from this, we usually find a single global minimum in the variational procedure 
(although the local minima tend to cluster in energy when more and more states are added). 
Thus, the full many-body ground state of the spin-boson model can be interpreted 
physically based on the optimal variational state, a path that we will follow
here.
We have also developed a new hierarchical algorithm for the optimization of 
the systematic variational state~(\ref{eq:GS}), see Appendix~\ref{Appendix}.

For the spin-boson model without any magnetic field along $\sigma_z$ and 
in absence of spontaneous symmetry breaking (which occurs for $\alpha>\alpha_c$), 
the system obeys a $\mathbb{Z}_2$ symmetry, so that the parameters for 
the ground state satisfy exactly
$p_m=-q_m$ and $f_k^{(m)}=-h_k^{(m)}$ for all $k$ and $m$.
This method was thoroughly tested for the Ohmic spin-boson model 
($s=1$)~\cite{Bera1,Bera2}, where extremely rapid convergence was 
established for a moderate number of coherent states $M_\mathrm{cs}\lesssim10$, 
unless one considers the deep Kondo regime where $\alpha\to1$.

\subsubsection{Full many-body wave function}

\begin{figure}[t]
\includegraphics[width=0.99\linewidth]{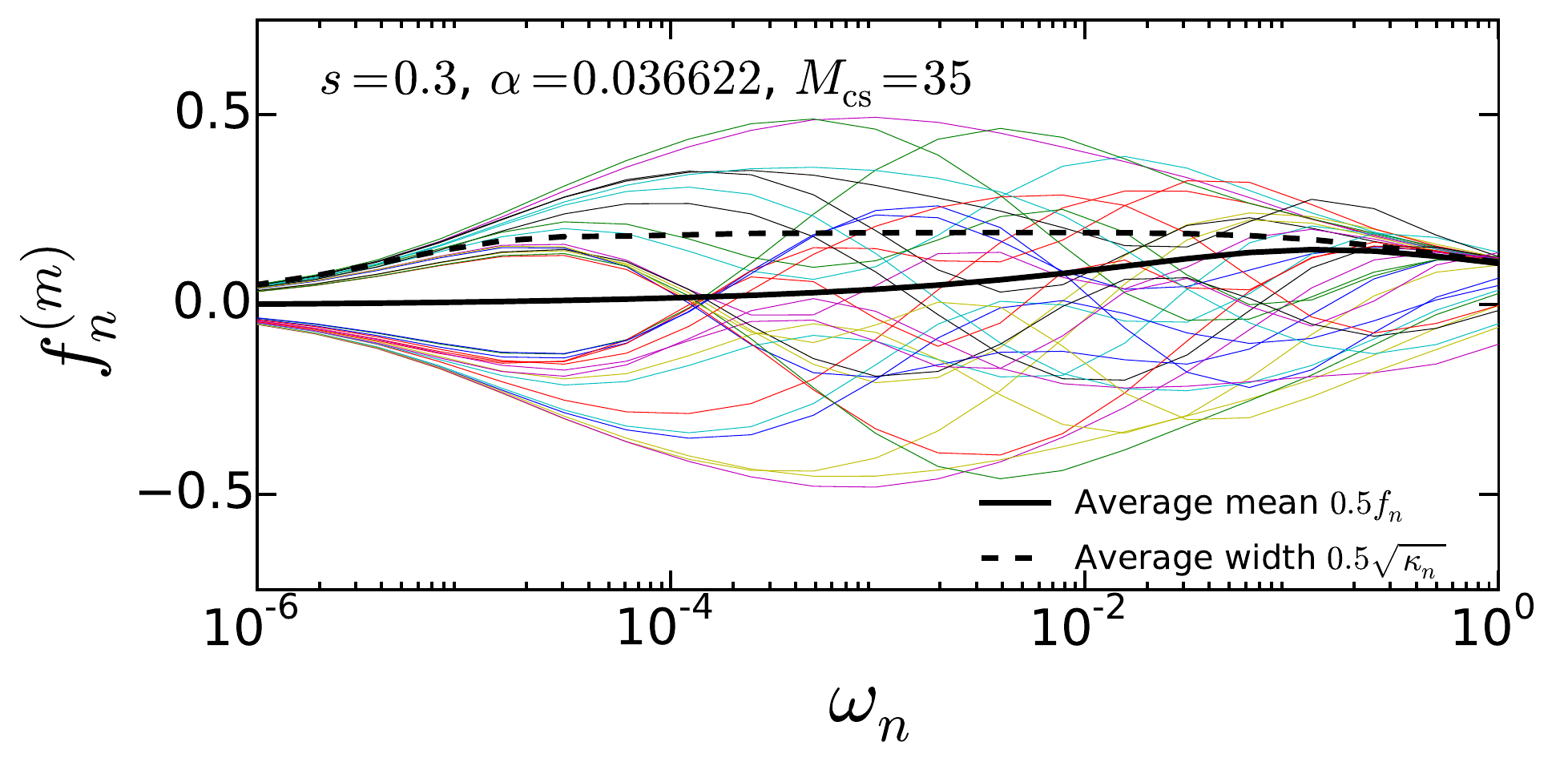}
\includegraphics[width=0.99\linewidth]{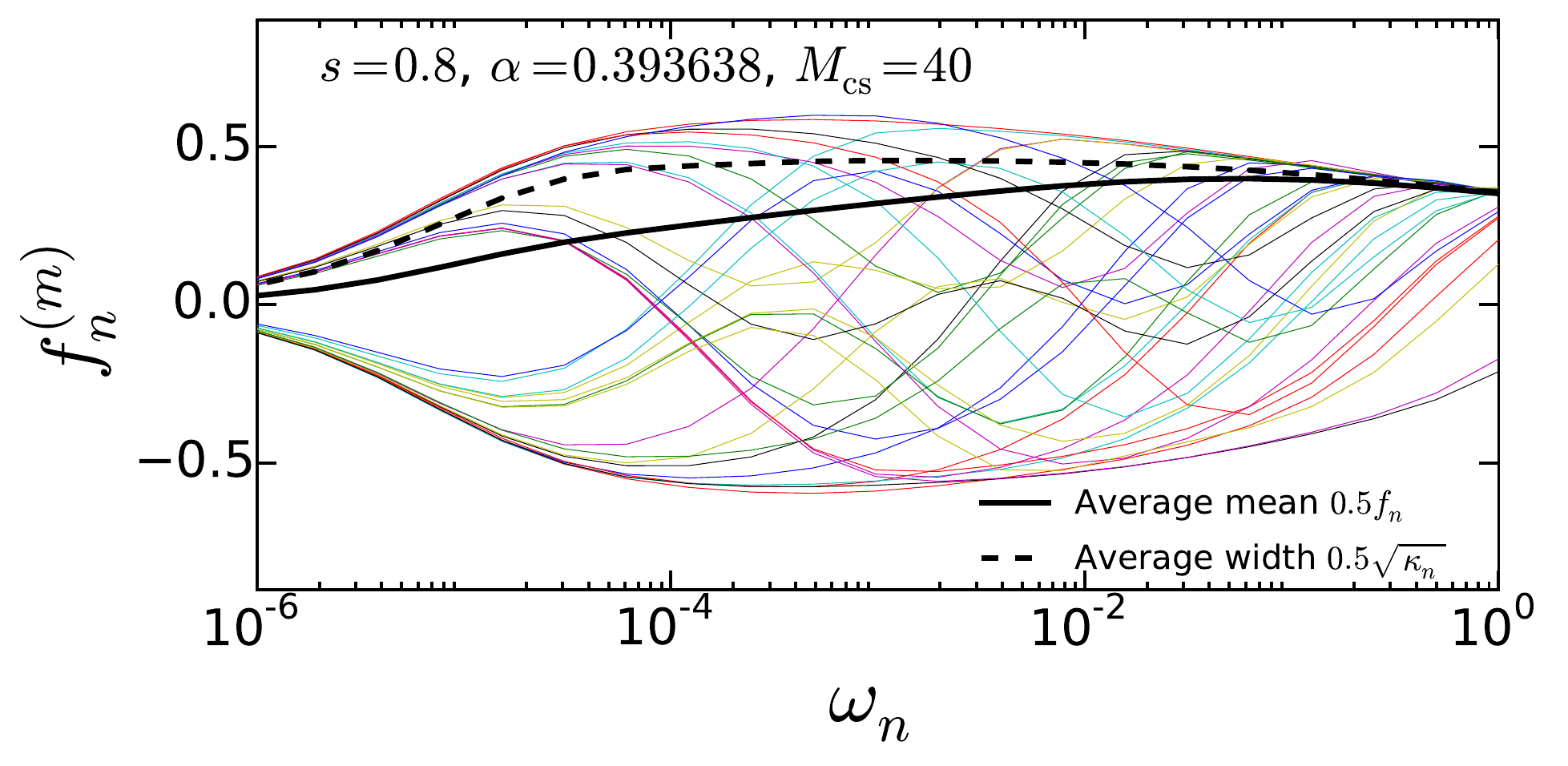}
\caption{(Color online) Nearly critical wave functions from the CSE for the case
$s=0.3$ (upper panel) and $s=0.8$ (lower panel), represented by the set of
displacements $f_n^{(m)}$ with $m=1\ldots M_\mathrm{cs}$ given by the thin full
lines. Thick full lines show the average mean displacement $0.5f_n$, and thick dashed 
lines the average width $0.5\sqrt{\kappa_n}$.
The critical regime is identified in the range $10^{-4}<\w_n<10^{-2}$ by a
constant plateau in $\kappa_n$, which reflects the clearly wide distribution of the
displacements associated to the classical-like configurations of the CSE. For frequencies
$\w\ll10^{-4}$, the wavefunction is no more critical and the displacements collapse 
onto a single curve, so that the distribution narrows, and $\kappa_n$ goes to zero.}
\label{FigWave}
\end{figure}

We show in Fig.~\ref{FigWave} typical wavefunctions obtained with the CSE near 
the quantum critical point (for two bath exponents $s=0.3$ and $s=0.8$). 
Here the set of displacements $f_n^{(m)}$ for each oscillator mode $a^\dagger_n$ 
is plotted versus the frequency $\omega_n$ of the mode, with $m=1\ldots M_\mathrm{cs}$ 
the index in the expansion (the corresponding weight $p_m$ are shown in 
the Appendix). In both plots, the critical domain lies roughly for frequencies 
in the range $10^{-4}<\w_n<10^{-2}$, which shows two striking observations.
First, the distribution of displacements looks very symmetric between positive
and negative values of the set of $f_n^{(m)}$, both in the critical regime, and
in the region of run-away flow $\w_n<10^{-4}$ at lower energy.
This symmetry is clearly not obeyed for the high energy modes near the cutoff.
Because the displacement operator directly couples to the order parameter $\sigma_z$ 
in Hamiltonian~(\ref{ham}), this symmetry nicely reflects the absence of spontaneous 
symmetry breaking at the critical point. This observation can be substantiated
mathematically by defining, from the star Hamiltonian~(\ref{star}), the average 
$f_n$ of the displacement fields in mode $n$ (see Sec.~\ref{Analytical} for thorough 
discussion):
\begin{eqnarray}
f_n &\equiv& \big<(\annd+\an)\sigma_z\big> \\
\nonumber
&=& 2\sum_{m,m'}^{M_{CS}} p_m p_{m'}\big<f^{\left( m\right)}|f^{\left(
m'\right)}\big>\left(f^{\left( m\right)}_n+f^{\left( m'\right)}_n\right).
\end{eqnarray}
The absence of spontaneous symmetry breaking, both at the critical point
and in the whole delocalized phase, translates in the fact that the average
value $f_n$ vanishes for $\w_n\to 0$. However, the set of displacements in the non-critical 
domain ($\omega_n<10^{-4}$) obey a trivial symmetry, as all displacements collapse on 
a single curve. In contrast, the displacements in the critical range
$10^{-4}<\w_n<10^{-2}$ keep fluctuating, showing a finite width of the distribution. 
This width $\kappa_n$ can be defined as follows:
\begin{eqnarray}
\kappa_n &\equiv& \big<(\annd+\an)^2\big>-1\\
\nonumber
&=& 2 \sum_{m,m'}^{M_{CS}}p_m p_{m'}\big<f^{\left(
m\right)}|f^{\left( m'\right)}\big>\left(f^{\left( m\right)}_n+f^{\left(
m'\right)}_n\right)^2.
\label{defkappa}
\end{eqnarray}
This plateau in $\kappa_n$ seen in the critical domain has for origin the
strong quantum fluctuations that take place at criticality, due to an order
parameter that is nearly but not quite localized. 
Alternatively, the width $\kappa_n$ can be interpreted as a squeezing 
parameter for the mode $a^\dagger_n$.

Having clarified the physics at play in the wave function itself, we will study 
these two coarse grained quantities $f_n$ (average) and $\kappa_n$ (width), which 
capture mathematically the distribution of classical configurations in 
the wave function. This study will rely not only on the CSE variational state, 
but also on VMPS calculations for benchmark, and on analytical field-theory 
calculations, which we present now.

\section{Analytical insights into various wave function properties}
\label{Analytical}

We establish in this section a set of exact analytical results for various
wave function properties, both in the non-critical and in the critical regimes.
The properties that we will consider concern the average displacement of
the bath oscillators, as well as their squeezing amplitude, which can be
interpreted as the variance of the oscillator displacements. These two
quantities thus give interesting information on the structure of the environmental 
wavefunction.

\subsection{Average displacement}
\subsubsection{General formula}
Owing to the linear coupling between $\sigma_z$ and the oscillator
displacement operator $(\akd+\ak)$, correlations are established between the
spin degree of freedom and its bosonic environment. Due to the
symmetry properties of Hamiltonian~(\ref{ham}),
the ground state wave function can be written generically as
$|\mr{GS}\big> = 
|\uparrow\big> |\Psi_\uparrow\big> -
|\downarrow\big> |\Psi_\downarrow\big>$, where
$|\Psi_\downarrow\big> = \hat{P} |\Psi_\uparrow\big>$,
with the parity operator $\hat{P}=\exp(i\pi\sum_k \akd \ak)$.
Thus, except for the trivial non-interacting case $\alpha=0$
where the environmental wave function is in the bare vacuum,
the qubit does not factorize from its environment. The manner in which 
correlations in $|\Psi_\uparrow\big>$ penetrate 
the bath states can be viewed equally as properties of a screening
cloud~\cite{Snyman1,Snyman2}. One goal of this paper is to illustrate
the behavior of this screening cloud in the sub-Ohmic model, both away 
from and at the quantum critical point. Since the environmental wave function 
$|\Psi_\uparrow\big>$ is a complicated object, the
simplest measure of the cloud resides in the average displacement $f_k$
that is obeyed by a given but arbitrary mode $\akd$ within this state. 
This quantity is defined as $f_k \equiv \big<(\akd+\ak)\s_z\big>$, where 
the average is taken with respect to the full many-body ground state $|\mr{GS}\big>$.
The average displacement $f_k$ gives thus information on how strong the order
parameter fluctuates at the energy scale $\omega_k$.

Now, we would like to show that this average displacement can be related exactly
to the spin-spin equilibrium correlation function, defined in imaginary 
time as:
\begin{equation}
\chi(\tau) = \big<\mr{GS}|T_\tau \frac{\sigma_z(\tau)}{2}\frac{\sigma_z(0)}{2}|\mr{GS}\big>,
\end{equation}
with $T_\tau$ the standard time-ordering operator, so that $T_\tau A(\tau) B(0)=
\theta(\tau)A(\tau)B(0)+\theta(-\tau)B(0)A(\tau)$. The imaginary-time
evolved operators read $A(\tau) = e^{H\tau}Ae^{-H\tau}$.
For the purpose of computing $f_k$, let us introduce the mixed correlation 
function between the spin and the displacement operator associated to
a given bosonic $k$-mode:
\begin{equation}
\label{mixed}
G_{z,k}(\tau)\equiv 
\big<\mr{GS}|T_\tau [\akd(\tau)+\ak(\tau)]\sigma_z(0)|\mr{GS}\big>,
\end{equation}
so that $f_k = G_{z,k}(0^+)$. Taking the time derivative in Eq.~(\ref{mixed}),
one gets the equations of motion:
\begin{equation}
\frac{\partial^2}{\partial\tau^2} G_{z,k}(\tau) = \w_k^2 G_{z,k}(\tau)
- 4 g_k \w_k \chi(\tau).
\end{equation}
Now, going to zero-temperature (but the formula below applies as well 
to finite temperature using discrete Matsubara frequencies), with 
$G(i\w) = \int_{-\infty}^{+\infty} \mr{d}t\; G(\tau) e^{i\w t}$,
one obtains the exact relation:
\begin{equation}
G_{z,k}(i\w) = \frac{4 g_k\w_k}{\w^2+\w_k^2} \chi(i\w).
\end{equation}
Going back to the time domain, one finds the connection between
the average displacement of the environmental wave function (the screening
cloud) and the local spin susceptibility:
\begin{equation}
\label{displacement}
f_k = \int \frac{\mr{d}\w}{2\pi} G_{z,k}(i\w) = 
4 g_k \w_k \int \frac{\mr{d}\w}{2\pi}
\frac{1}{\w^2+\w_k^2} \chi(i\w).
\end{equation}
From this equation, previous knowledge obtained for spin dynamics of 
the sub-Ohmic model~\cite{Bulla,Tong,Freyn} will allow us to make exact 
predictions for the average displacement characterizing the screening
cloud.

\subsubsection{Asymptotic behavior of the average displacement}

A change of variable in Eq.~(\ref{displacement}) gives
\begin{equation}
f_k =
4 g_k \int \frac{\mr{d}x}{2\pi}
\frac{1}{x^2+1} \chi(i\w_k x),
\end{equation}
so that the small-momentum behavior of $f_k$ is determined by the low-energy 
scaling of the spin-spin correlation function~\cite{Bulla,Tong,Freyn}, which reads
$\chi(i\w)\simeq 1/(m_R+B_s|\w|^s)$, with $B_s=4\alpha\w_c^{1-s}
\int dx x^{s-1}/(1+x^2)$. Here $m_R$ is the 
renormalized mass, which is finite in the delocalized phase ($\alpha<\alpha_c$) 
and vanishes at the quantum critical point. Thus, two scaling laws 
are established in the limit $k\to0$:
\begin{eqnarray}
\label{fNonCritical}
f_k &\simeq& \frac{2g_k}{m_R} \;\; \mr{for} \;\; \alpha<\alpha_c,\\
f_k &\simeq& \frac{4 A_s}{B_s} \frac{g_k}{|\w_k|^s} \;\; \mr{for} \;\; \alpha=\alpha_c,
\end{eqnarray}
where $A_s = \int (dx/2\pi) x^{-s}/(1+x^2)$.
Let us now specialize to the case of the Wilson energy discretization on the grid
$\w_n = \w_c \Lambda^{-n}$, in which case $\w_k$ is replaced by 
$\xi_n\propto \Lambda^{-n}\propto \w_n$ and $g_k$ by $\gamma_n/\sqrt{\pi}\propto 
\Lambda^{-n(s+1)/2} \propto \w_n^{(s+1)/2}$.
We thus find the following low-energy scaling laws of the average displacement 
for the modes obeying the Wilson energy discretization:
\begin{eqnarray}
f_n &\propto& \w_n^{(1+s)/2} \;\; \mr{for} \;\; \alpha<\alpha_c,\\
f_n &\propto& \w_n^{(1-s)/2} \;\; \mr{for} \;\; \alpha=\alpha_c.
\end{eqnarray}
The non-critical modes thus follow a different and faster power law than 
the critical ones, a result that we shall confirm from our numerics in 
Sec.~\ref{Comparison}. In fact, our low frequency analysis allows us to
extract the exact prefactor of the critical
average displacement. At $\alpha=\alpha_c$, we find:
\begin{equation}
f_n=\frac{\sqrt{2}\left(s+2\right)^s\left(1-\Lambda^{-\left(s+1\right)}\right)^{s+\frac{1}{2}} \tan\frac{\pi s}{2}}
{\pi\sqrt{\alpha} \omega_c^{\frac{1-s}{2}}\left(s+1\right)^{s+\frac{1}{2}}\left(1-\Lambda^{-\left(s+2\right)}\right)^s}
 \; \omega_n^{\frac{1-s}{2}}.
\label{fCritical}
\end{equation}
The prefactor is clearly non-universal, as a dependence in the frequency
cutoff $\omega_c$ is present.

\subsection{Average width (squeezing amplitude)}
\subsubsection{General formula}
Generalizing the previous results, we define the average intramode squeezing amplitude 
as $\kappa_k \equiv \big<(\akd+\ak)^2\big>-1$, such that it is exactly zero for a 
vacuum state. Following the previous methodology, we introduce the intermode
Green's function of the displacement field of the bosonic modes:
\begin{equation}
\label{inter}
G_{k,q}(\tau)\equiv 
\big<\mr{GS}|T_\tau [\akd(\tau)+\ak(\tau)][\aqd(0)+\aq(0)]|\mr{GS}\big>.
\end{equation}
Applying the time-derivative twice provides exact equations of motion, which 
lead to the following formula in the Matsubara domain:
\begin{equation}
G_{k,q}(i\w)\equiv 
G_{k}^0(i\w)\delta_{k,q} + g_k g_q G_{k}^0(i\w)G_{q}^0(i\w) \chi(i\w),
\end{equation}
where $G_{k}^0(i\w)= 2 \w_k/(\w^2+\w_k^2)$.
This gives the exact equation relating the average squeezing parameter to the
dynamical spin-spin susceptibility:
\begin{equation}
\kappa_k = \int \frac{\mr{d}\w}{2\pi} G_{k,k}(i\w) - 1
= 4 g_k^2 \w_k^2 \int \frac{\mr{d}\w}{2\pi}
\frac{1}{(\w^2+\w_k^2)^2} \chi(i\w).
\label{squeezing}
\end{equation}
Again, knowledge of the spin dynamics will give information on
the average squeezing parameter for the ground state wave function.

\subsubsection{Asymptotic behavior of the average squeezing}

Similar to our analysis of the average displacement, a change of 
variable in Eq.~(\ref{squeezing}) gives
\begin{equation}
\kappa_k =
\frac{4 g_k^2}{\w_k} \int \frac{\mr{d}x}{2\pi}
\frac{1}{(x^2+1)^2} \chi(i\w_k x),
\end{equation}
resulting in the following low-energy leading order behavior of the average squeezing 
amplitude:
\begin{eqnarray}
\kappa_k &\simeq& \frac{g_k^2}{m\w_k} \;\; \mr{for} \;\; \alpha<\alpha_c,\\
\kappa_k &\simeq& \frac{4 C_s}{B_s} \frac{g_k^2}{|\w_k|^{1+s}} \;\; \mr{for}
\;\; \alpha=\alpha_c,
\end{eqnarray}
where $C_s = \int (dx/2\pi) x^{-s}/(1+x^2)^2$.
In the case of the Wilson energy discretization on the grid
$\w_n = \w_c \Lambda^{-n}$, we get the explicit scaling laws for the
average squeezing amplitude:
\begin{eqnarray}
\kappa_n &\propto&\w_n^{s} \;\; \mr{for} \;\; \alpha<\alpha_c,\\
\nonumber
\kappa_n&=& \mathrm{const.} \;\; \mr{for} \;\; \alpha=\alpha_c.
\end{eqnarray}
We find a constant and universal (cutoff independent) value of $\kappa_n$ at 
the quantum critical point, as a precise computation of the constant value
for $\alpha=\alpha_c$ reads: 
\begin{equation}
\kappa_n=\frac{\left(s+2\right)^{s+1}\left(1-\Lambda^{-\left(s+1\right)}\right)^{s+2}\tan{\frac{\pi s}{2}}}
{\pi\left(s+1\right)^{s+1}\left(1-\Lambda^{-\left(s+2\right)}\right)^{s+1}}.
\label{kappaCritical}
\end{equation}
Since the average displacement $f_n$ vanishes at low energy, this means that the 
distribution of displacements of the critical wave function is very broad, reflecting 
the strong fluctuations of the order parameter at the quantum critical point. 
We stress that $\kappa_n$, defined as (\ref{defkappa}), strictly vanishes in the 
continuum limit $\Lambda\to1$, but that it remains finite when integrated over a
logarithmic energy mesh.

\begin{figure}[t]
\includegraphics[width=0.92\linewidth]{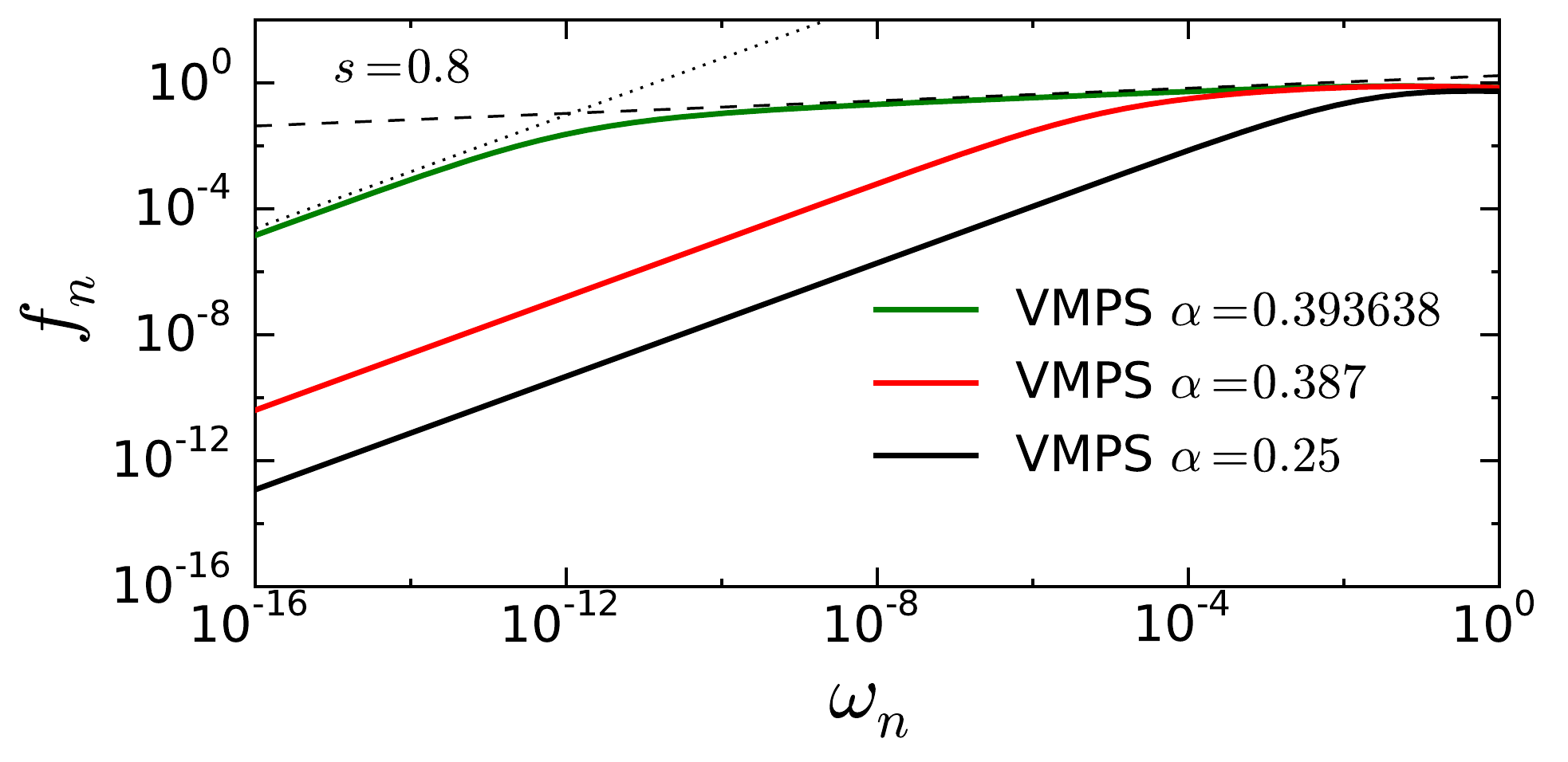}
\includegraphics[width=0.99\linewidth]{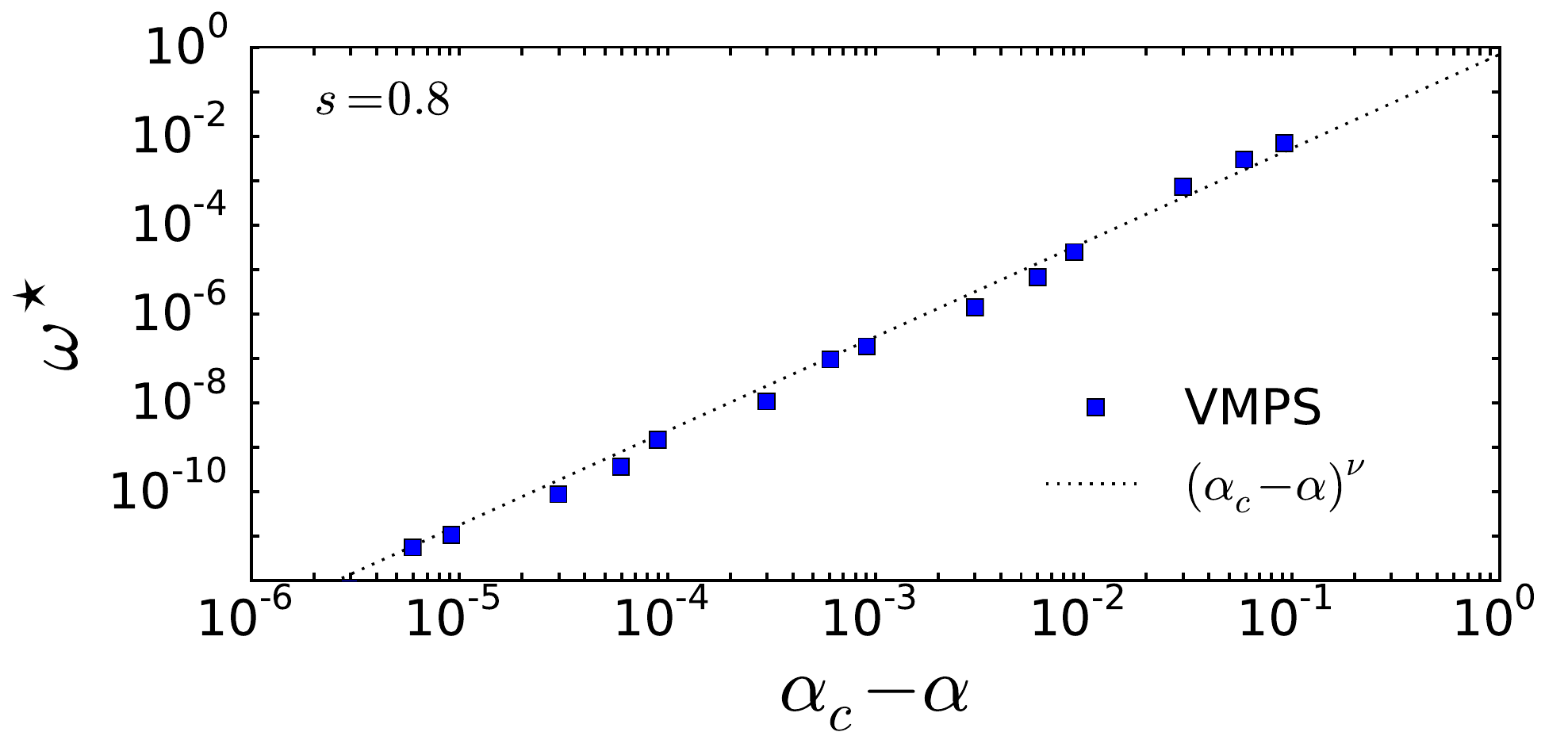}
\caption{(Color online) The upper panel shows the average displacement 
$f_n=\big< (\annd + a_n)\sigma_z \big>$ from the VMPS calculation at $s=0.8$, for three 
values of $\alpha=0.25, 0.387, 0.393638$ (the last value is very close to 
the quantum critical interaction strength $\alpha_c$).
The dotted line denotes the non-critical scaling
$f_n\propto \w_n^{(1+s)/2}$ for $\w\ll\w^\star$, while the dashed line 
indicates the expected critical behavior $f_n\propto \w_n^{(1-s)/2}$
for $\w^\star\ll\w\ll\w_c$.  The crossover scale $\w^\star$ between the two
scaling behaviors is shown in the lower panel for a large selection of $\alpha$ 
values, allowing us to extract the correlation length exponent $\nu\simeq0.47$ for 
$s=0.8$. This value is quite different from the mean-field result 
$\nu_\mathrm{MF}=1/s=1.25$, because the system lies below its upper 
critical dimension~\cite{Bulla,Tong}.}
\label{VMPSonly1}
\end{figure}
\begin{figure}[t]
\includegraphics[width=0.89\linewidth]{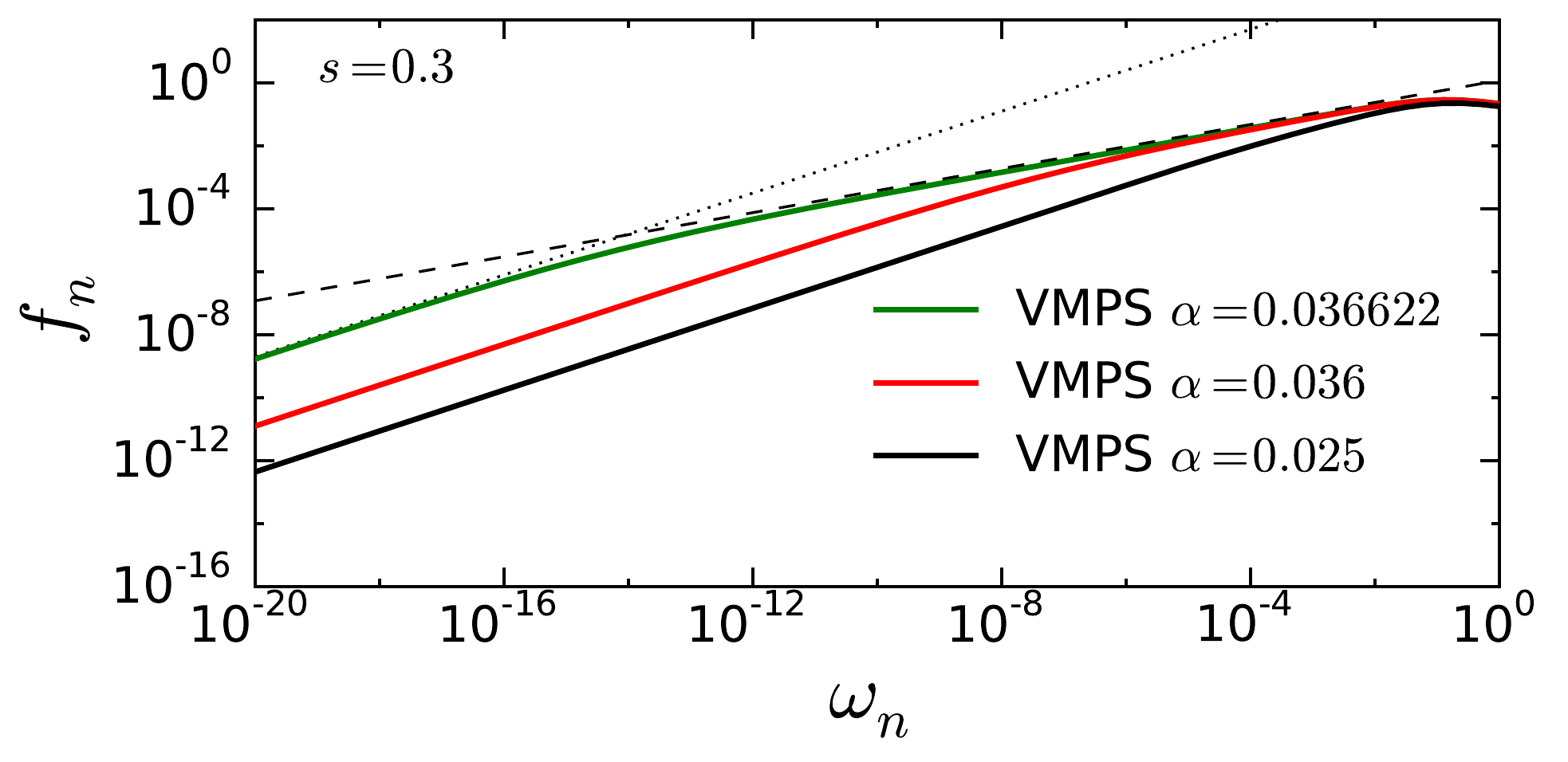}
\includegraphics[width=0.90\linewidth]{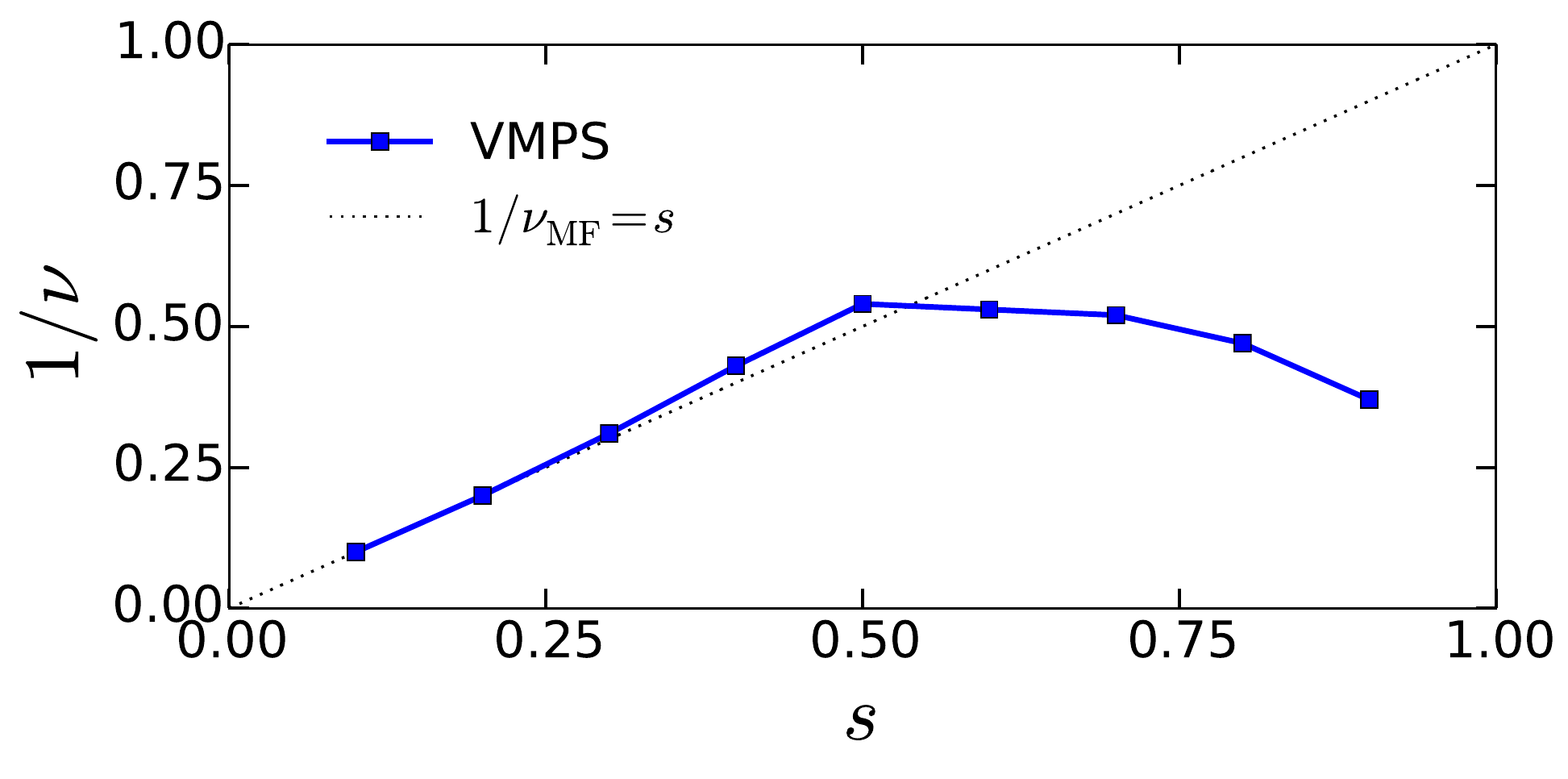}
\caption{(Color online) The upper panel shows, similarly to
Fig.~\ref{VMPSonly1}, the average displacement, but now for $s=0.3$, 
with $\alpha=0.0326, 0.036, 0.036622$ (the last value is 
very close to the quantum critical point). The dotted 
line shows the non-critical scaling $f_n\propto \w_n^{(1+s)/2}$,
while the dashed line indicates the expected critical
quantum behavior $f_n\propto \w_n^{(1-s)/2}$. 
The lower panel shows the extracted correlation length exponent 
$\nu$ for various values of $s$, which assumes the mean-field prediction
$\nu_\mathrm{MF}=1/s$ only for $0<s<1/2$.}
\label{VMPSonly2}
\end{figure}

\section{Numerical results}
\label{Comparison}

\subsection{General scaling behavior}

We start by presenting general VMPS calculations, allowing us to outline 
the scaling behavior and the quantum criticality of the sub-Ohmic
spin-boson model.
We shall consider two different values of the bath 
spectral density throughout the paper, $s=0.3$ and $s=0.8$. The former corresponds to the case 
where the quantum phase transition is of mean-field type, while the latter 
case is associated to an interacting fixed point~\cite{Bulla,Tong,Rieger,Anders}. 
We stress beforehand that both the average displacement $f_n$ and average squeezing 
amplitude $\kappa_n$ are exactly 
related to the dynamical susceptibility from Eqs.~(\ref{displacement}) 
and (\ref{squeezing}), so that their scaling behavior as a function of 
momenta, both in the non-critical and critical regimes, is determined by 
a trivial $s$-dependent exponent. 

However, non-trivial exponents in the interacting case $0.5<s<1$ will show 
up in the $\alpha$-dependence of the correlation length $\xi$ that is defined
by the spatial extent up to which quantum critical fluctuations penetrates
within the bath states. More precisely, the correlation length is given by an inverse energy
$\xi=1/\w^\star$, where $\w^\star$ is such that quantum critical behavior is 
established for $\w^\star \ll \w_k \ll \w_c$ (this regime sets in only if $\alpha$ 
is quite close to $\alpha_c$).
This correlation length behaves as $\xi \propto |\alpha_c-\alpha|^{-\nu}$, 
with the exponent $\nu_\mathrm{MF}=1/s$ in the mean field regime $0<s<1/2$. This can be 
gathered from the low-energy behavior $\chi(i\w)\simeq 1/(m_R+B_s|\w|^s)$
and the absence of singular vertex corrections at mean-field level, giving
the renormalized mass $m_R\propto \alpha_c-\alpha$. However, $\nu$ assumes 
non-trivial values given by a classical long-range Ising model~\cite{Bulla,Tong} for 
the interacting regime $1/2<s<1$. This behavior is illustrated in the lower 
panels of Fig.~\ref{VMPSonly1} and Fig.~\ref{VMPSonly2}.
Thus, both the average displacement and average squeezing amplitude (not shown here) encode 
non-trivial exponents for $1/2<s<1$, but only due to the divergent correlation 
length $\xi=1/\w^\star$.
These observations can be also summed up by scaling laws:
\begin{eqnarray}
f_n &=& \w_n^{(1-s)/2} F(\w_n/\w^\star),\\
\kappa_n &=& K(\w_n/\w^\star),
\end{eqnarray}
with $F(x),K(x)\propto 1$ for $x\gg1$, and $F(x),K(x)\propto x^{s}$ for $x\ll1$.
This general scaling behavior of the average displacement is illustrated in the upper panel 
of Fig.~\ref{VMPSonly1} for $s=0.8$, and in the upper panel of
Fig.~\ref{VMPSonly2} for $s=0.3$.
We find indeed that our VMPS data exhibits the expected non-critical
and critical scaling laws, respectively $f_n\propto \w_n^{(1+s)/2}$
for $\alpha\ll\alpha_c$ (dotted line) and $f_n\propto \w_n^{(1-s)/2}$ 
for $\alpha=\alpha_c$ (dashed line).
We now turn to a more detailed analysis, with a comparison to our analytical
predictions, and with the numerics from the coherent state expansion. 

\begin{figure}[t]
\includegraphics[width=0.99\linewidth]{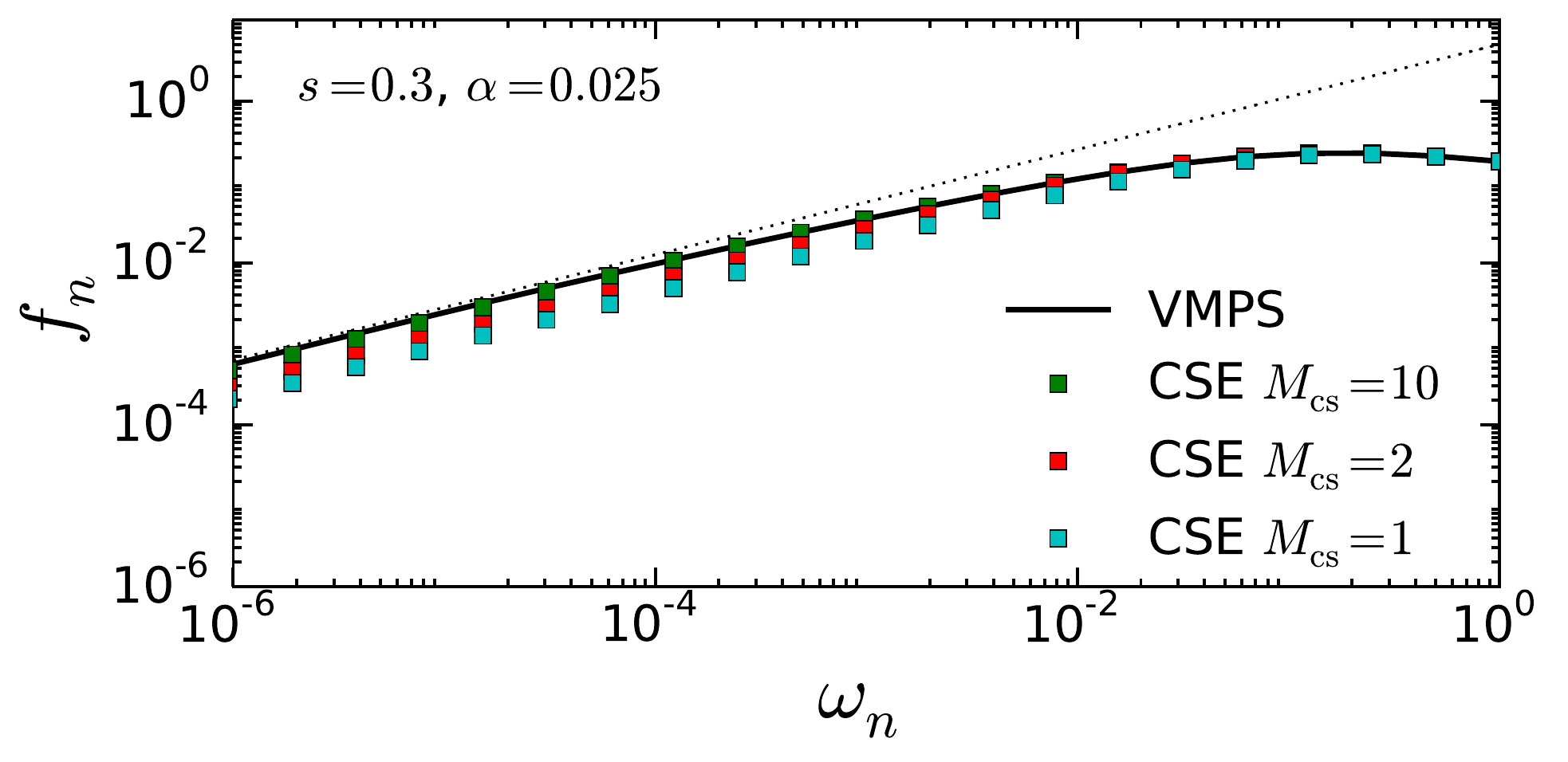}
\includegraphics[width=0.99\linewidth]{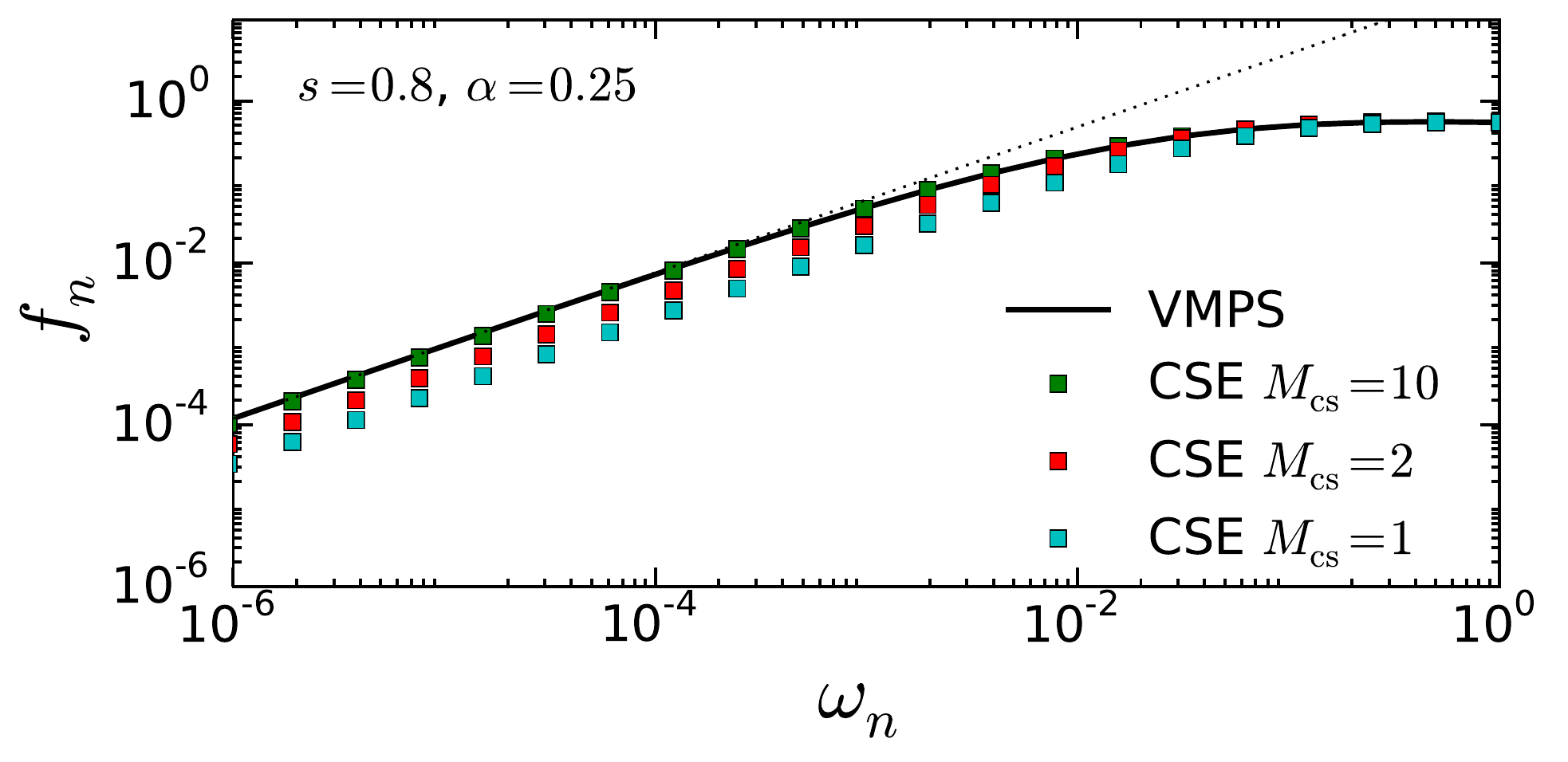}
\caption{(Color online) Average displacement $f_n$ of mode $\annd$ in the non-critical 
regime ($\alpha\ll\alpha_c$) for two values of the bath spectra, $s=0.3$ (top panel, 
with $\alpha=0.025$) and $s=0.8$ (bottom panel, with $\alpha=0.25$). 
The full black line denotes the fully converged VMPS results, while the colored symbols show the CSE
at increasing number of coherent states, $M_\mr{cs}=1,2,10$ (bottom to top). 
A dotted line denotes the expected $f_n\propto \w_n^{(1+s)/2}$ behavior in the 
non-critical regime.}
\label{FigDisPre}
\end{figure}

\subsection{Non-critical regime}
Let us now investigate the non-critical regime, which is established either for $\alpha\ll\alpha_c$
at all frequencies, or for $\alpha\simeq\alpha_c$ but for $\w\ll\w^\star$.
Focusing first on the average displacement, we consider in Fig.~\ref{FigDisPre} the
two cases $s=0.3$ and $s=0.8$ for values of $\alpha$ that are sufficiently away
from $\alpha_c$ so that critical behavior is not triggered. 
The comparison between the fully converged VMPS data and CSE at increasing number
$M_\mr{cs}$ of coherent states shows that the CSE converges very quickly 
in this simplest non-critical regime. In addition, the CSE captures the exact 
leading behavior of the average displacement, $f_n\propto \w_n^{(1+s)/2}$, already for 
$M_\mr{cs}=1$ (the so-called Silbey-Harris theory~\cite{Emery,Silbey,ChinSubOhmic}), 
since the variational equation gives $f_k = (g_k/2)/(\w_k+\Delta_R)\propto g_k/\Delta_R$ 
for $k\to0$, in agreement with the exact result~(\ref{fNonCritical}). Note that the 
quantum critical scaling $f_n\propto \w_n^{(1-s)/2}$ is
not apparent in this plot, because the $\alpha$ value is too far away
from $\alpha_c$.

\begin{figure}[t]
\includegraphics[width=0.99\linewidth]{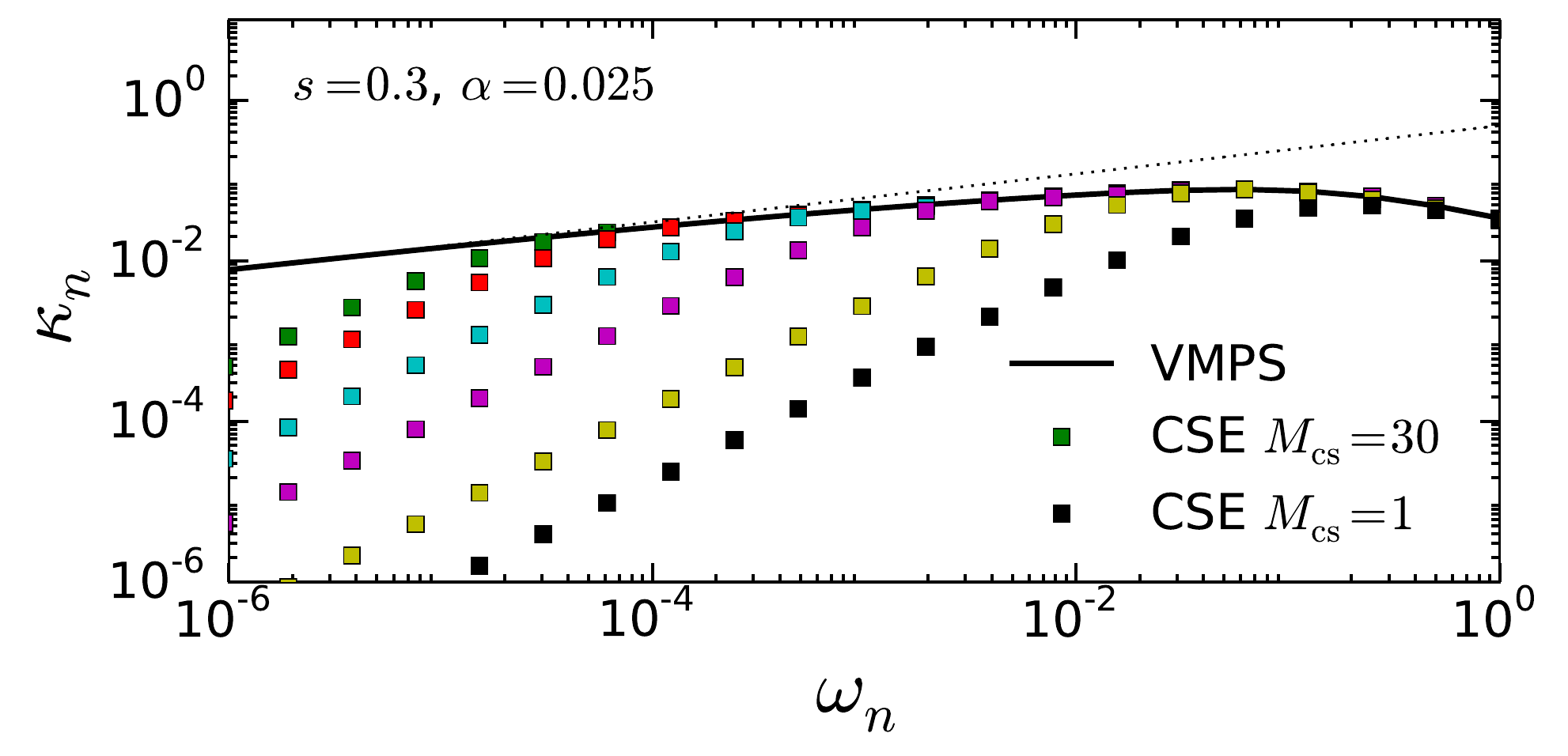}
\includegraphics[width=0.99\linewidth]{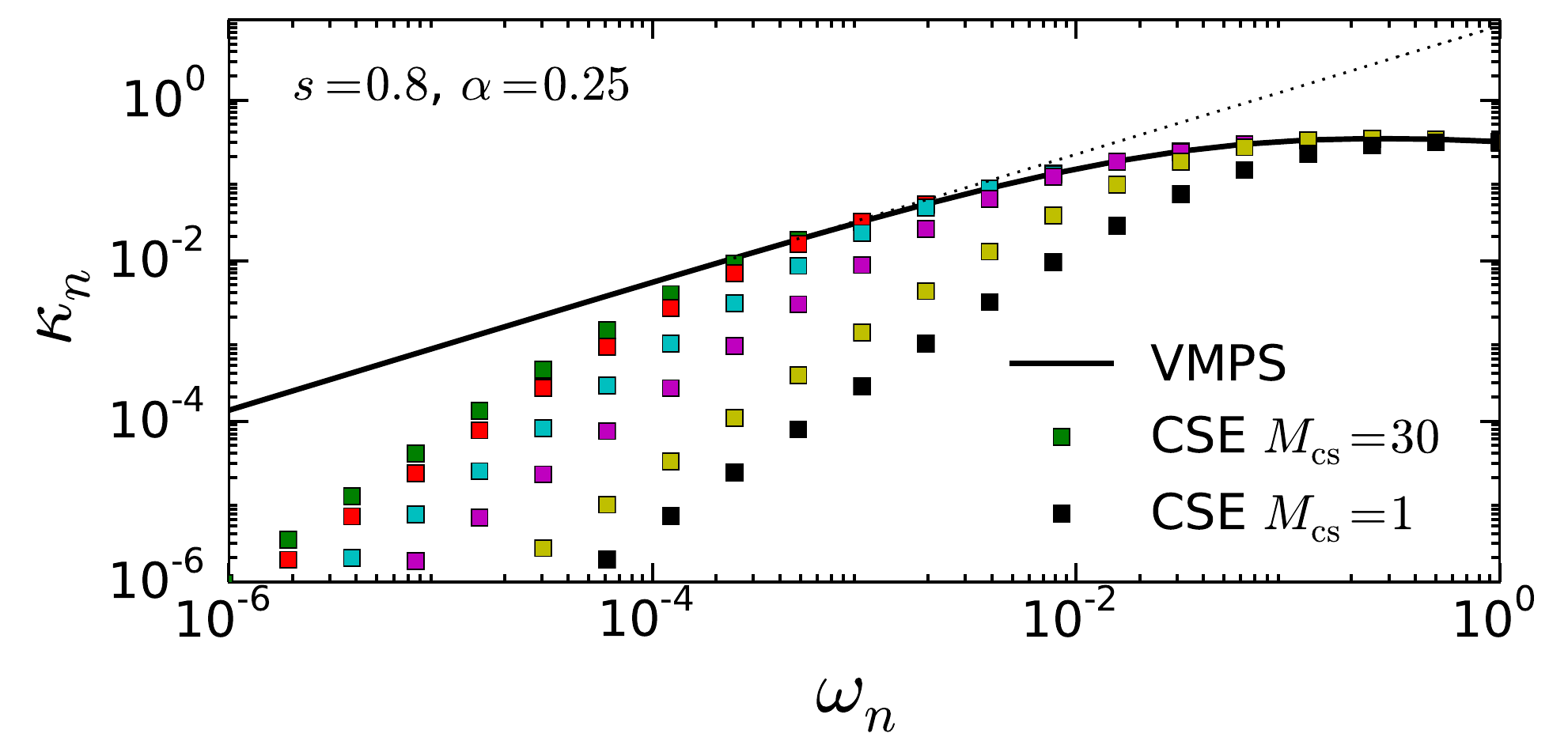}
\caption{(Color online) Average squeezing amplitude $\kappa_n$ of mode $\annd$ in the
non-critical regime ($\alpha\ll\alpha_c$) for two values of the bath spectra, $s=0.3$ 
(top panel, with $\alpha=0.025$) and $s=0.8$ (bottom panel, with $\alpha=0.25$). 
The full black line denotes the fully converged VMPS results, while the colored symbols show the CSE
at increasing number of coherent states, $M_\mr{cs}=1,2,5,10,20,30$ (bottom to
top).
A dotted line denotes the expected $\kappa_n\propto \w_n^{s}$ behavior in 
the non-critical regime.}
\label{FigSquPre}
\end{figure}

Turning to the average squeezing amplitude, we find excellent agreement of our
converged CSE results to the expected non-critical scaling behavior
$\kappa_n\propto \w_n^{s}$, see Fig.~\ref{FigSquPre}. 
However, we observe a much slower convergence of the CSE for the average squeezing
amplitude as compared to the computation of the average displacements in Fig.~\ref{FigDisPre}, 
especially regarding the low-energy modes. This behavior can be understood from the Silbey-Harris
theory at $M_{cs}=1$, which predicts incorrectly $\kappa_n=(f_n)^2\propto 
\w_n^{1+s}$ instead of the exact non-critical scaling $\kappa_n\propto \w_n^{s}$.
This disagreement is not fully a surprise, because the Silbey-Harris theory is based
on a single coherent state, and is tailored to address at best the 
displacement and not necessarily the squeezing amplitude. As a matter of
fact, one can prove from the explicit form of the displacements~\cite{Bera2} 
at arbitrary $M_\mr{cs}$ values that the incorrect scaling behavior $\kappa_n \propto 
\w_n^{1+s}$ at vanishing $\w_n$ is found for any finite value of $M_\mr{cs}$, 
which is also clear from Fig.~\ref{FigSquPre}. Only in the strict limit $M_\mr{cs}\to\infty$ is
the correct non-critical scaling obeyed down to zero energy. Nevertheless, if one focuses
on a reasonable energy range (typically a few decades), the correct non-critical scaling 
behavior is well captured for both the average displacement and the average squeezing 
amplitude in our CSE computations. This analysis illustrates the general fact that
systematic variational calculations may lead to the rapid convergence of some physical 
observables, but not of others. This problem is particularly severe near quantum
critical points, because the deviations concern asymptotically low-energy modes,
which occupy a tiny fraction of the total ground state energy.

\subsection{Critical regime}

We now consider the quantum critical point, where the dissipation strength
$\alpha=\alpha_c$ is such that the correlation length $\xi=1/\w^\star$
diverges. In practice we fine tune $\alpha_c-\alpha$ to more than 7 digits
so that $\xi$ is larger than $10^{10}$, as can be seen from the VMPS data
of Fig.~\ref{VMPSonly1}. The coherent state expansion offers, alternatively,
a more pictorial view of the quantum critical wave function, which can be
fully represented by a set of classical-like displacement configurations,
as shown previously in Fig.~\ref{FigWave}.

While the average critical displacement $f_n\propto \w_n^{(1-s)/2}$ vanishes 
(with the expected exponent) at low energy, we showed analytically
Eq.~(\ref{kappaCritical}) that the average squeezing amplitude
$\kappa_n=\big<(\annd+\an)^2\big>-1$ is constant at the quantum critical point. 
Thus $\kappa_n$ can be viewed as the average fluctuation of the displacements
within the many-body wave function. Therefore we conclude that $\kappa_n\gg (f_n)^2$ 
at the quantum critical point, which reflects the strong fluctuations of the order
parameter. This expected physical picture is very clear in Fig.~\ref{FigWave}:
in the intermediate energy range $10^{-4}<\w<10^{-2}$,
the distribution of displacements is nearly symmetric around zero, and thus
almost vanishes on average (this behavior is more pronounced for $s=0.3$ than
for $s=0.8$, because the average displacement vanishes as $\omega_n^{(1-s)/2}$).
In contrast, the width of the distribution of displacements has roughly a constant value
in the critical domain.
Away from the critical domain, namely for very low frequencies $\w\ll\w^\star$,
the distribution of the classical-like configurations becomes very narrow as all
displacements collapse onto the same curve. Thus the average squeezing amplitude
should vanish, with the non-critical scaling behavior $\kappa_n\propto\w_n^s$.
However, due to the finite size of the coherent state basis set used here, we
find for this computation the different behavior $\kappa_n\propto\w_n^{1+s}$
as discussed previously.

\begin{figure}[t]
\includegraphics[width=0.99\linewidth]{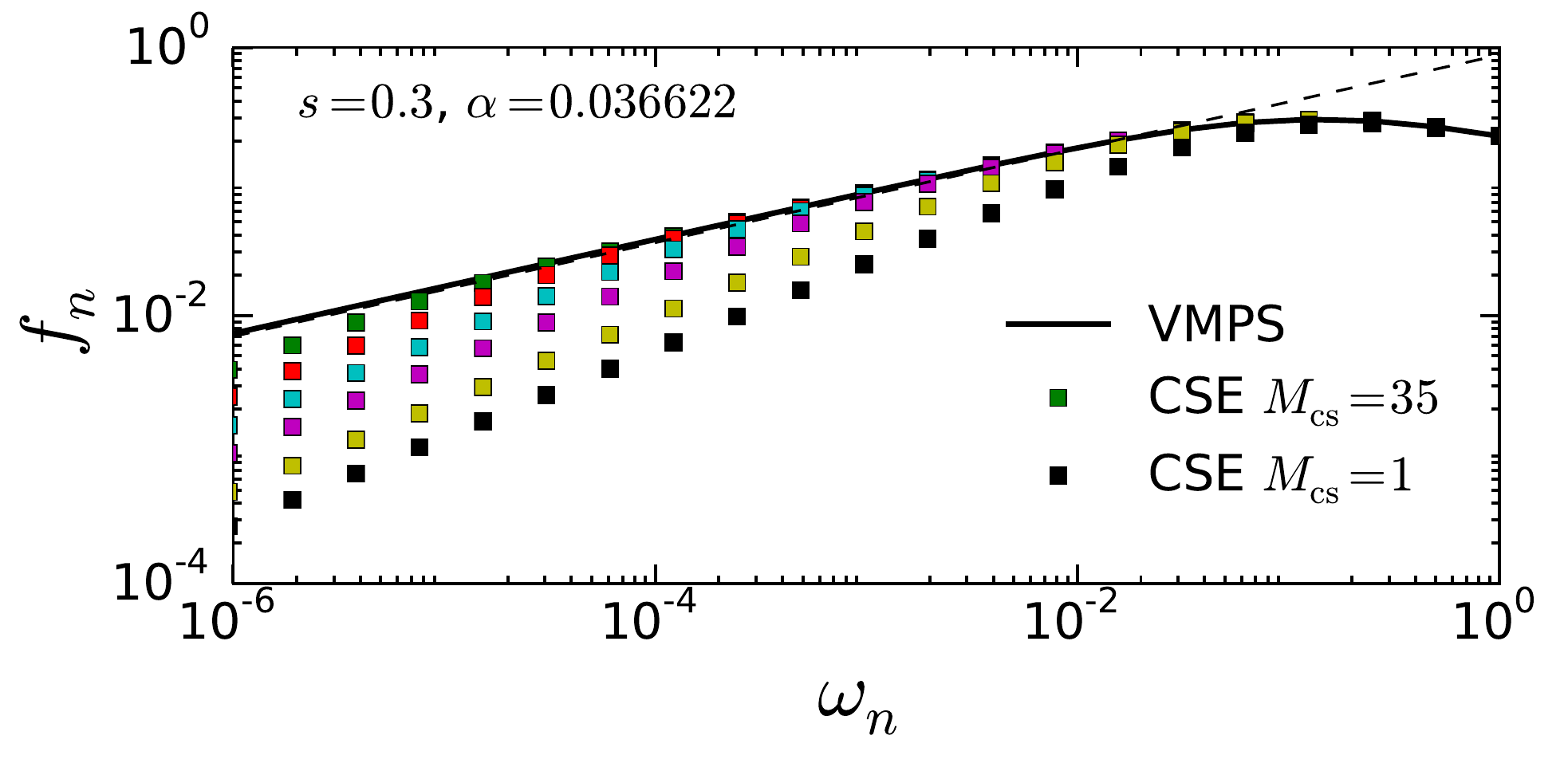}
\includegraphics[width=0.99\linewidth]{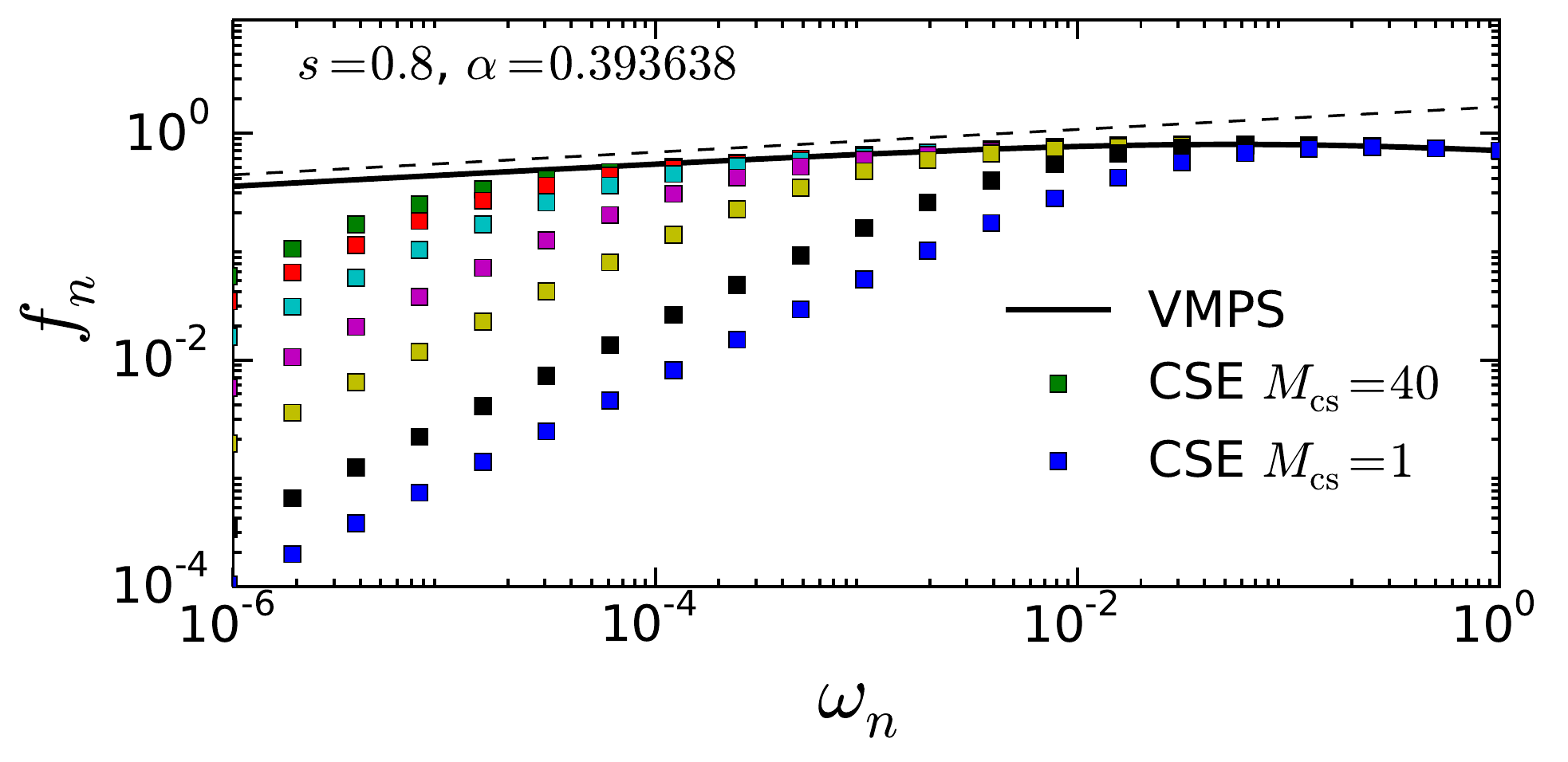}
\caption{(Color online) Average displacement $f_n$ of mode $\annd$ at the critical 
point ($\alpha=\alpha_c$) for two values of the bath spectra, $s=0.3$ (top panel, 
with $\alpha=0.03662$) and $s=0.8$ (bottom panel, with $\alpha=0.393638$).
The full black line denotes the fully converged VMPS results, while the colored symbols 
show the CSE at increasing number of coherent states, $M_\mr{cs}=1,2,5,10,20,35$ 
(bottom to top) for $s=0.3$ and $M_\mr{cs}=1,2,5,10,20,30,40$ for $s=0.8$. 
A dashed line denotes the expected $f_n \simeq F_s \w_n^{(1-s)/2}$ 
scaling behavior in the critical regime, including the analytical prefactor $F_s$ given
in Eq.~(\ref{fCritical}).}
\label{FigDisCri}
\end{figure}

\begin{figure}[t]
\includegraphics[width=0.99\linewidth]{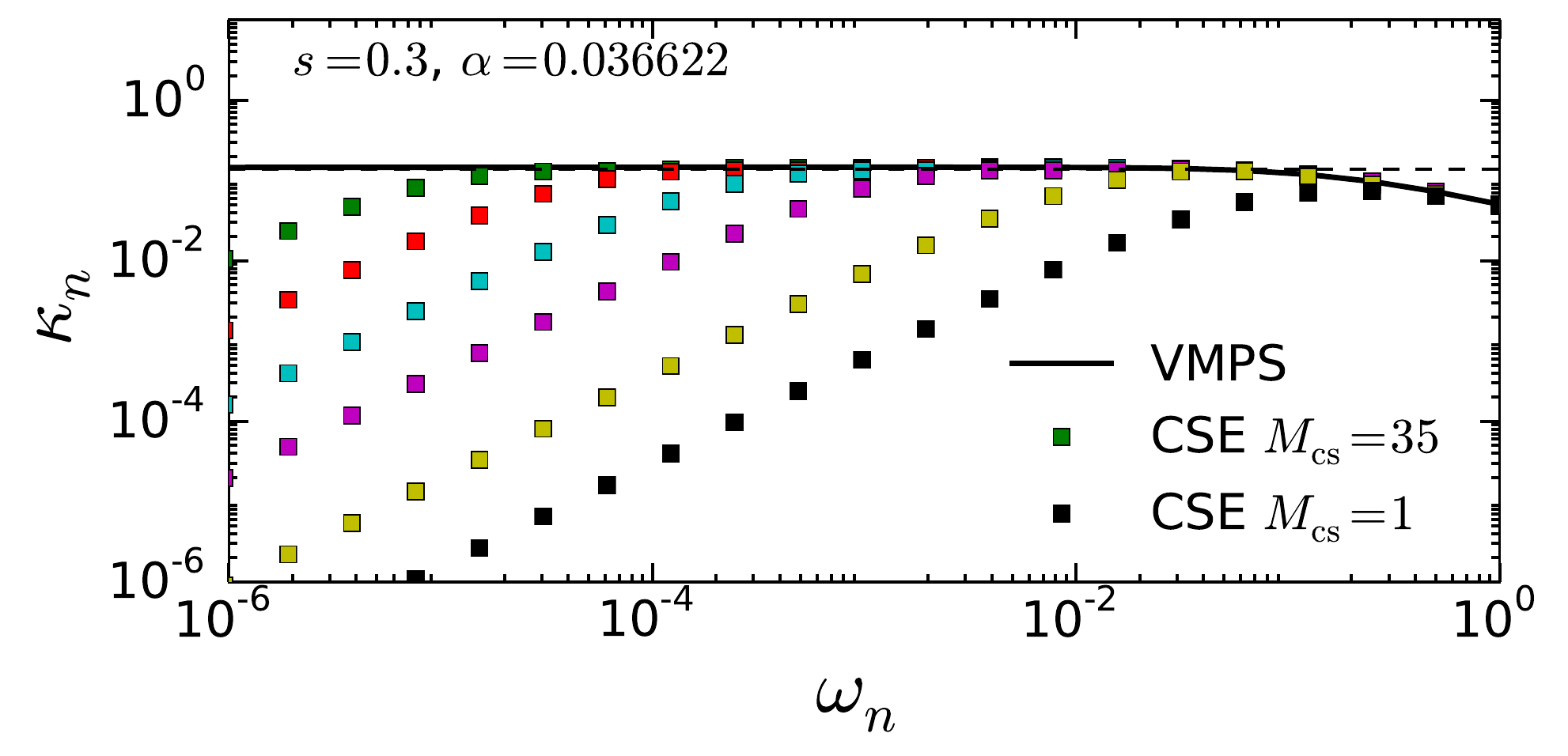}
\includegraphics[width=0.99\linewidth]{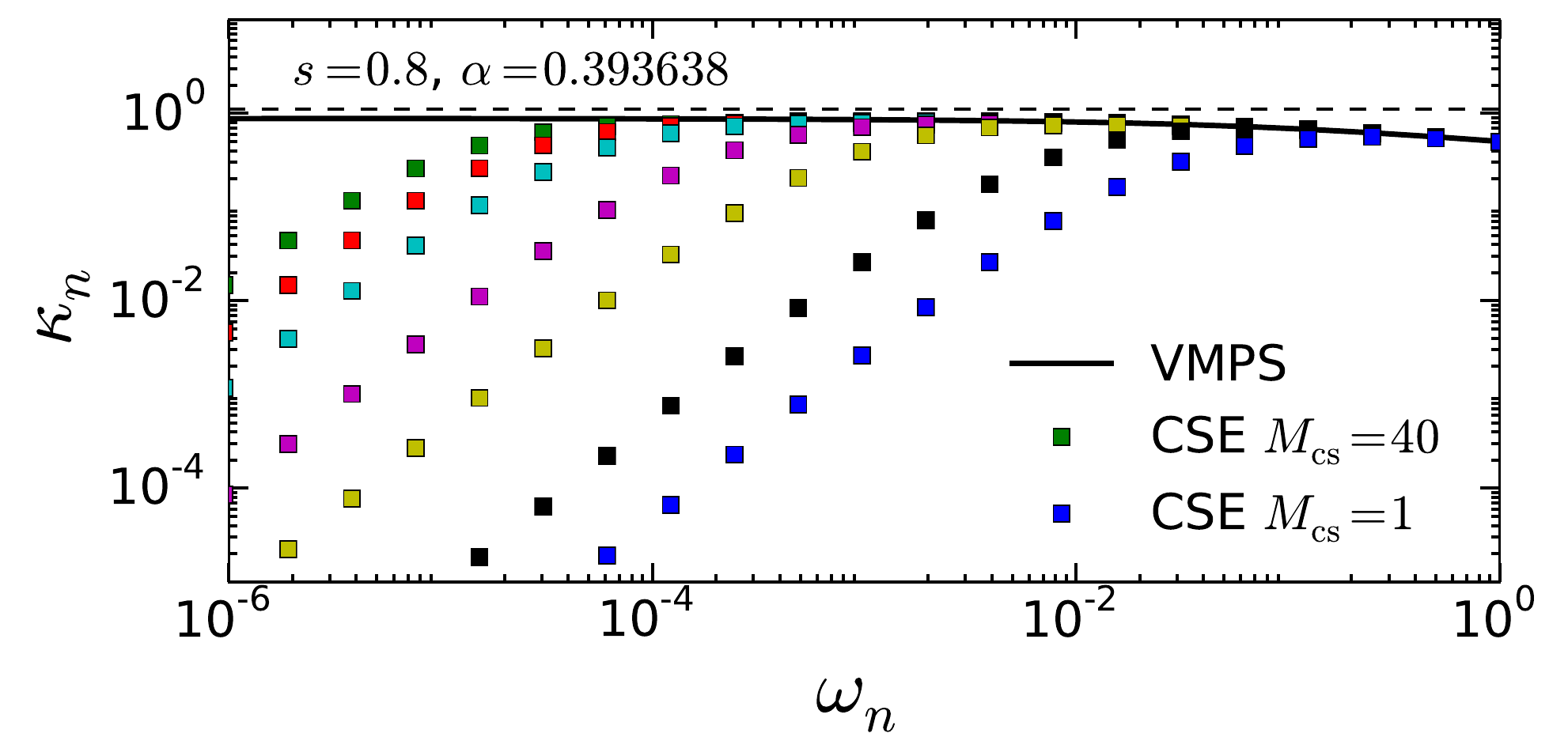}
\caption{(Color online) Average squeezing amplitude $\kappa_n$ of mode $\annd$ at the critical 
point ($\alpha=\alpha_c$) for two values of the bath spectra, $s=0.3$ (top panel, 
with $\alpha=0.03662$) and $s=0.8$ (bottom panel, with $\alpha=0.393638$).
The full black line denotes the fully converged VMPS results, while the colored symbols show 
the CSE at increasing number of coherent states, $M_\mr{cs}=1,2,5,10,20,35$ (bottom 
to top) for $s=0.3$ and $M_\mr{cs}=1,2,5,10,20,30,40$ for $s=0.8$.
A dashed line denotes the expected universal constant value of $\kappa_n$ given by
Eq.~(\ref{kappaCritical}) in the critical regime.}
\label{FigSquCri}
\end{figure}

Let us finally check in more detail the precise scaling behavior of the
critical average displacement in Fig.~\ref{FigDisCri}.
Again we find excellent convergence of the CSE to the VMPS curves, and
we are able to match quantitatively the expected scaling law
Eq.~(\ref{fCritical}), including the analytic prefactor $F_s$ in front
of the power law $\w_n^{(1-s)/2}$. Due to
the construction of the CSE based on coherent states, one sees again that
any truncation to finite $M_\mr{cs}$ produces an incorrect scaling $f_n\propto
\w_n^{(1+s)/2}$ at vanishing energy. But the correct power law
is typically obeyed on several decades for a moderate numerical effort.
The same type of behavior is also found in the critical average squeezing 
amplitude $\kappa_n$, which shows the expected constant plateau, see
Fig.~\ref{FigSquCri}, and that matches the analytical prediction of
Eq.~(\ref{kappaCritical}) nicely.
We have assessed the general prediction of the power-law dependance of the
critical average displacement $f_n\propto \w_n^{(1-s)/2}$ by fitting the low
energy tails of our converged data for a wide selection of the bath exponent 
$s$ in the range $0<s<1$. We found that the critical exponent $(1-s)/2$ is very 
well obeyed, both in the mean-field and interacting regimes, with an accuracy
of a few percent. This reflects the peculiarity of the spin-boson model,
which does not present anomalous exponents in the spin-spin correlation
function~\cite{Bulla,Tong,Freyn}, even below its upper critical dimension.


\section{Conclusion}
We have investigated physical properties of ground-state wave functions in a
simple model of quantum criticality, the sub-Ohmic spin-boson Hamiltonian.
For this purpose, a combination of variational matrix product states and
an extensive coherent state expansion have been performed and compared
very precisely. The coherent state approach allows a direct representation
of many-body wave functions in terms of a collection of classical-like
trajectories associated to a set of displacements. Focusing on the quantum
critical regime, the wave function displays a nearly symmetric distribution of 
displacements at low energy. However its width, related to a squeezing amplitude 
of the low-energy modes defined on a logarithmic energy interval, remains finite 
with a universal value. 
This behavior strikingly reflects the wide fluctuations of the order parameter 
at the quantum critical point in absence of spontaneous symmetry breaking, in 
analogy with strong statistical fluctuations near classical phase transitions.
Detailed analytical predictions have been made using exact field theory
results, which match very well all the obtained numerical data, both in
the non-critical and critical regimes. Similar analysis should be possible
for various extensions of dissipative impurity model, such as the two-bath
case~\cite{Guo,Bruognolo}, which presents new classes of interacting fixed
points. It should be applicable also to fermionic models, both with
impurities or with bulk interactions, using a similar decomposition of
the many-body wavefunction in terms of a distribution of one- or two-body 
phase shifts~\cite{Snyman3}.

\acknowledgments We thank H. Baranger, N. Roch and I. Snyman for useful discussions.
SF, AN, AWC, and ZBC acknowledge funding from the CNRS PICS contract ``StrongCircQED'' 
(No. 191148). AN and ZBC are also supported by The University of Manchester and the EPSRC. 
AWC acknowledges support from the Winton Programme for the Physics of
Sustainability. BB and JvD are supported by the DFG through the Excellence Cluster ``Nanosystems 
Initiative Munich'', SFB/TR 12, SFB 631.

\appendix
\section{Hierarchical algorithm for the CSE}
\label{Appendix}
We present here a new algorithm for finding the many-body ground state~(\ref{eq:GS})
of the spin-boson model~(\ref{ham}), which radically improves the methodology developed
previously in Refs.~\cite{Bera1,Bera2}, allowing us to incorporate a large number
$M_\mr{cs}$ of coherent states.
This new scheme, devised to optimize efficiently the energy functional, is only
based on fast local minimization routines. Indeed, while global minimization 
routines such as simulated annealing can give the most reliable estimates, 
they do not scale favorably in the case of a large number of variational 
parameters. However, blind application of local routines, for instance L-BFGS
or conjugate gradients~\cite{CG}, do not guarantee convergence to the lowest 
energy minimum. Hence, physical insight must be used as a guide to implement a fast
and reliable local optimization method. 

\begin{figure}[h]
\includegraphics[width=0.99\linewidth]{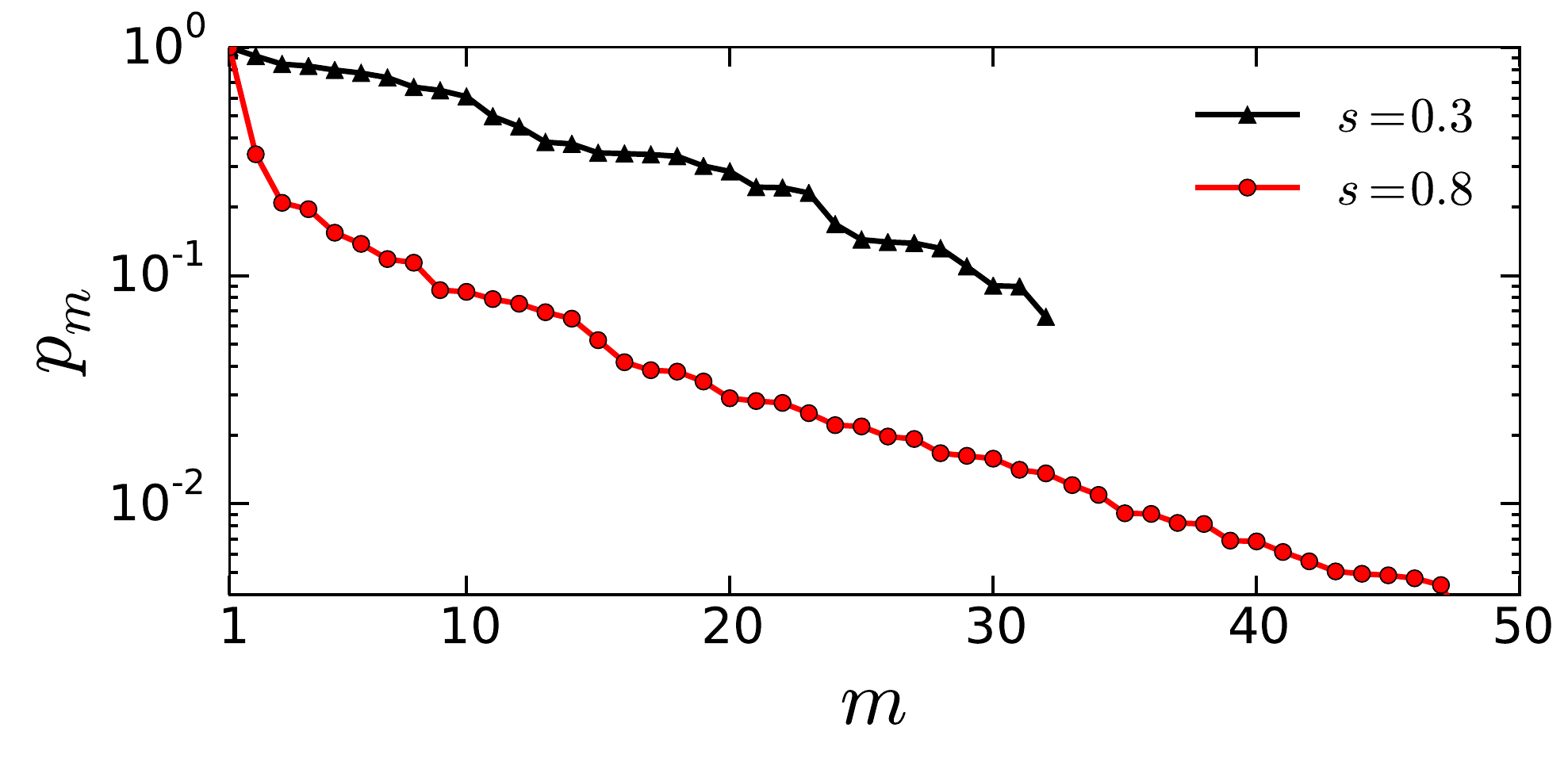}
\caption{(Color online) Weights $p_m$ of the coherent state $|f^{(m)}\big>$ in 
the coherent state expansion~(\ref{eq:GS}), as a function of index $m$, 
for $s=0.3$ (triangles) and $s=0.8$ (circles), with the same parameters as in
Fig.~\ref{FigWave}. The fast exponential decay of the weights illustrates the 
hierarchical structure of the CSE.}
\label{FigWeights}
\end{figure}

Here, we use the fact that the coherent state decomposition~(\ref{eq:GS}) is an 
expansion that displays a hierarchical structure. Indeed, our simulations demonstrate 
that the weight $p_M$ of a newly added coherent state is typically smaller 
than the majority of the weights $p_m$ of the preceding states.
This hierarchical structure is clearly apparent in Fig.~\ref{FigWeights}.

This feature is exploited as follows in our numerical implementation. The
algorithm starts with the solution for a single coherent state (the so-called
Silbey-Harris Ansatz) with $M_\mr{cs}=1$, which is reliably obtained by a local
routine, providing a first estimate of $f_k^{(1)}$. Then the energy is minimized
for $M_\mr{cs}=2$ with two coherent states, using the previously determined
$f_k^{(1)}$ as an initial guess, $f_k^{(2)}=0$, and $p_2=p_1/2$. Both displacements 
(and their corresponding weights) are then optimized together. The algorithm continues
in the same manner by increasing $M_\mr{cs}$ by one unit at a time, and using
the previous displacements and weights as an initial guess for the next
minimization stage.
For completeness, we give below all the required analytical expressions used
in our simulations.

\subsection{Explicit form of the energy functional}
We focus here on the case of $\mathbb{Z}_2$ symmetry, so that the averaged
Hamiltonian from the systematic variational state~(\ref{eq:GS}) reads:
\begin{eqnarray}
\big<H\big> &=& -\Delta\sum_{n,m=1}^{M_\mr{cs}} p_n p_m \big<f^{(n)}|-f^{(m)}\big>\\
\nonumber
&&+\sum_{n,m=1}^{M_\mr{cs}} p_n p_m \big<f^{(n)}|f^{(m)}\big> 
\sum_q 2\w_q f_q^{(n)} f_q^{(m)}\\
\nonumber
&&-\sum_{n,m=1}^{M_\mr{cs}} p_n p_m \big<f^{(n)}|f^{(m)}\big> 
\sum_q g_q\left(f_q^{(n)}+f_q^{(m)}\right).
\end{eqnarray}
The overlaps obey the usual coherent state algebra (all displacements
$f_q^{(n)}$ and weights $p_n$ are
real in the ground state), namely $\big<f^{(n)}|f^{(m)}\big> =
e^{-(1/2)\sum_q(f_q^{(n)}-f_q^{(m)})^2}$.
The minimization is performed on the energy
$E=\big<H\big>/\mathcal{N}$
with the norm $\mathcal{N}=\big<\mr{GS}|\mr{GS}\big>=
2\sum_{n,m=1}^{M_\mr{cs}} p_n p_m \big<f^{(n)}|f^{(m)}\big>$.

\subsection{Energy gradients}
Standard optimization routines gain a huge computing advantage by using
an explicit expression for the gradient of the function to be minimized.
We thus provide here the gradients with respect to the weight $p_M$ and 
displacement $f_k^{(M)}$:
\begin{widetext}
\begin{eqnarray}
\frac{\partial E}{\partial p_M} &=& \frac{2}{\mathcal{N}}
\sum_{n=1}^{M_\mr{cs}} p_n \Bigg\{
\! -\Delta \big<f^{(n)}|-f^{(M)}\big>
+\big<f^{(n)}|f^{(M)}\big>
\Big[ \sum_q \left( 2\w_q f_q^{(n)} f_q^{(M)} -
g_q \left(f_q^{(n)}+f_q^{(M)}\right)\right)-2 E\Big]\Bigg\},\\
\frac{\partial E}{\partial f_k^{(M)}} &=& \frac{2p_M}{\mathcal{N}}
\sum_{n=1}^{M_\mr{cs}} p_n \Bigg\{
\! \Delta \big<f^{(n)}|-f^{(M)}\big> \Big(f_k^{(n)}+f_k^{(M)}\Big)
+ \big<f^{(n)}|f^{(M)}\big> 
\Big(2\w_k f_k^{(n)}-g_k\Big)\\
\nonumber
&& \hspace{2cm} 
+ \big<f^{(n)}|f^{(M)}\big> \Big(f_k^{(n)}-f_k^{(M)}\Big)
\Big[
\sum_q \left( 2\w_q f_q^{(n)} f_q^{(M)} -
g_q \left(f_q^{(n)}+f_q^{(M)}\right)\right)-2 E
\Big]\Bigg\}.
\end{eqnarray}
\end{widetext}

\end{document}